\documentclass[prd,aps,twocolumn,a4paper,floatfix,nofootinbib]{revtex4-2}

\usepackage{graphicx}
\usepackage{dcolumn}
\usepackage{bm}
\usepackage{amsmath}
\usepackage{graphicx}
\usepackage{array}
\usepackage{booktabs}
\usepackage{xcolor}
\usepackage{amssymb}
\usepackage{pifont}
\usepackage{multirow}
\usepackage{diagbox}
\usepackage{comment}
\usepackage{xcolor}
\usepackage{float}
\usepackage{enumerate}
\usepackage{makecell}
\usepackage{booktabs}
\usepackage[normalem]{ulem}
\usepackage[utf8]{inputenc}
\usepackage{hyperref}
\allowdisplaybreaks[1]
\usepackage{units}
\usepackage{enumerate}
\usepackage{enumitem}
\hypersetup{
    colorlinks = true,
    linkcolor = {blue},
    citecolor = {blue},
    urlcolor = {blue},
    linkbordercolor = {white},
    }
\graphicspath{ {./Figures/} }

\def\d{\mathrm{d}}

\def\mF{\mathcal{F}}

\newcommand{\tmatch}{{t_\mathrm{match}}}
\newcommand{\keff}{{k_l^\mathrm{eff}}}
\newcommand{\dt}{\Delta t_{22}}
\newcommand{\MSun}{M_\odot}

\begin{document}

\def\vfivet{{\texttt{SEOBNRv5THM}}}
\def\vfourt{{\texttt{SEOBNRv4T}}}
\def\vfivehm{{\texttt{SEOBNRv5HM}}}
\def\vfiveHM{{\texttt{SEOBNRv5HM}}}
\def\nrtidalvthree{{\texttt{NRTidalv3}}}
\def\nrtidal{{\texttt{NRTidal}}}
\def\nrtidalvtwo{{\texttt{NRTidalv2}}}
\def\nrtvthree{{\texttt{NRTidalv3}}}
\def\imrphenomdnrtidaltwo{{\texttt{IMRPhenomD\_NRTidalv2}}}
\def\imrphenomdnrtidalthree{{\texttt{IMRPhenomD\_NRTidalv3}}}
\def\imrphenomxasnrtidaltwo{{\texttt{IMRPhenomXAS\_NRTidalv2}}}
\def\xasnrtidalvthree{{\texttt{IMRPhenomXAS\_NRTidalv3}}}
\def\imrphenomxpnrtidalthree{{\texttt{IMRPhenomXP\_NRTidalv3}}}
\def\imrphenomxpnrtidaltwo{{\texttt{IMRPhenomXP\_NRTidalv2}}}
\def\imrphenompvtwonrtidaltwo{{\texttt{IMRPhenomPv2\_NRTidalv2}}}
\def\vfourT{{\texttt{SEOBNRv4T}}}
\def\vfiveT{{\texttt{SEOBNRv5THM}}}
\def\lalsuite{{\texttt{LALSuite}}}
\def\seobnrvfiverom{{\texttt{SEOBNRv5\_ROM}}}
\def\seobnrvfourromnrtidalvtwo{{\texttt{SEOBNRv4\_ROM\_NRTidalv2}}}
\def\seobnrvfiveromnrtidalvtwo{{\texttt{SEOBNRv5\_ROM\_NRTidalv2}}}
\def\seobnrvfiveromnrtidalvthree{{\texttt{SEOBNRv5\_ROM\_NRTidalv3}}}
\def\teob{{\texttt{TEOBResumS}}}
\def\teobg{{\texttt{TEOBResumS-Giotto}}}

\title{Modeling matter(s) in \texttt{SEOBNRv5THM}: Generating fast and accurate effective-one-body waveforms for spin-aligned binary neutron stars}

\author{Marcus \surname{Haberland}$^{1}$}\email{marcus.haberland@aei.mpg.de}
\author{Alessandra \surname{Buonanno}$^{1,2}$}\email{alessandra.buonanno@aei.mpg.de}
\author{Jan \surname{Steinhoff}$^{1}$}\email{jan.steinhoff@aei.mpg.de}

\affiliation{${}^1$Max Planck Institute for Gravitational Physics (Albert Einstein Institute), Am M\"uhlenberg 1, Potsdam 14476, Germany}
\affiliation{${}^2$Department of Physics, University of Maryland, College Park, MD 20742, USA}

\date{\today}

\begin{abstract}
    We present \texttt{SEOBNRv5THM}, an accurate and fast gravitational-waveform model for quasi-circular, spinning, non-precessing binary neutron stars (BNS) within the effective-one-body (EOB) formalism. It builds on the binary--black-hole approximant \texttt{SEOBNRv5HM}\ and, compared to its predecessor \texttt{SEOBNRv4T}, it i) incorporates recent high-order post-Newtonian results in the inspiral, including higher-order adiabatic tidal contributions, spin-induced multipoles and dynamical tides for spin-aligned neutron stars, ii) includes the gravitational modes $(\ell, |m|)=(2,2),(3,3),(2,1),(4,4),(5,5),(3,2),$ and $(4,3)$, iii) has a time of merger calibrated to BNS numerical-relativity (NR) simulations, iv) accurately models the pre-merger $(2,2)$ mode through a novel phenomenological ansatz, and v) is 100 to 1000 times faster than its predecessor model for BNS systems with total mass $M \geq 2\, M_\odot$. Thus, \texttt{SEOBNRv5THM} can be used in Bayesian parameter estimation, which we perform for two BNS events observed by the LIGO-Virgo Collaboration, GW170817 and GW190425. The model accurately reproduces BAM and SACRA NR waveforms with errors comparable to or lower than the intrinsic NR uncertainty. We validate the model against the other state-of-the-art BNS waveform models \texttt{NRTidalv3}\ and \texttt{TEOBResumS}\ and find differences only for highly spinning and highly tidally deformable BNS, where there is no NR coverage and the models employ different spin prescriptions. Our model serves as a foundation for the development of subsequent \texttt{SEOBNR}\ waveform models with matter that incorporate further effects, such as spin-precession and eccentricity, to be employed for upcoming observing runs of the LIGO-Virgo-KAGRA Collaboration and future facilities on the ground.
\end{abstract}


\maketitle

\section{\label{section: introduction}Introduction}

The field of gravitational-wave (GW) astronomy has undergone rapid expansion since the initial detection of GWs from a binary--black-hole (BBH) merger in 2015~\cite{LIGOScientific:2016aoc}, and the first detection of a binary--neutron-star (BNS) merger only two years later, in the landmark event GW170817~\cite{LIGOScientific:2017vwq}. Among the 90 GW detections of the LIGO-Virgo-KAGRA (LVK) Collaboration up to and including the third observing run (O3)~\cite{GWTC3,GWTC2.1}, GW170817 was the first to be observable in both GWs and electromagnetically across the entire spectrum from radio to gamma rays, yielding novel insights into astrophysics, tests of General Relativity and cosmology~\cite{LIGOScientific:2017vwq, LIGOScientific:2017zic, LIGOScientific:2017ync}, as well as establishing the era of multi-messenger astronomy with GWs. The fourth observing run (O4) of the LVK Collaboration is currently ongoing, during which one binary system containing a neutron star (NS) has been observed so far~\cite{GW230529}.

The Bayesian analysis framework~\cite{Veitch:2014wba,Thrane:2018qnx} facilitates the extraction of physical parameters, such as masses, spins, and the macroscopic description of matter, like the tidal deformability of a NS, from GW signals. To this end, observed data is compared with theoretical predictions derived from the Einstein field equations.

Over the next decades, the observed merger rates of BNS will significantly increase, up to multiple thousands per year (see, e.g., Refs.~\cite{Iacovelli:2022,3G_Science_Book}), as existing detector facilities become more sensitive, and as new facilities on the ground, such as the Einstein Telescope~\cite{Punturo:2010zz}, and Cosmic Explorer~\cite{Reitze:2019iox,Evans:2021gyd}, are constructed.

Two challenges have to be overcome to extract reliable and exhaustive knowledge. As detector sensitivities increase, higher signal-to-noise ratios demand more accurate waveform models to avoid parameter-estimation biases~\cite{Samajdar:2018,Gamba:2021prd,Kunert:2022,Kunert:2024}. With a new plethora of observational data, we will additionally be challenged to drastically reduce waveform generation speed for the parameter-estimation (PE), as the current Bayesian analysis approach will become computationally infeasible. PE codes based on machine-learning methods, notably neural posterior estimation, are also under continuous development to speed up inference studies~\cite{Dax:2021tsq,Dax:2022pxd,Dax:2024}.

Importantly, by enhancing the speed and precision of existing waveform models, significant insights into NSs can be obtained, going beyond the progress already obtained with GW170817~\cite{LIGOScientific:2017vwq, Dietrich:2020efo}. The structure of the NS is primarily described by a macroscopic equation of state (EoS) of the neutron-rich supranuclear-dense matter~\cite{Lattimer:2012nd, Ozel:2016oaf}. This EoS serves to supplement the Tolman-Oppenheimer-Volkoff equations, in order to make predictions about the NS's interior.
Given that the matter in the core of a NS is a few times denser than the nuclear saturation density, it cannot be constrained by experiments in earth-based accelerators and reactors due to its extreme pressure and internal energy. Consequently, there are various realistic theoretical and phenomenological models in nuclear physics (see, e.g., Refs.~\cite{Chin:1974sa, Serot:1997xg, Huth:2021bsp, Alford:2022bpp, Lattimer:2012nd, Lattimer:2021emm, Burgio:2021vgk, Zhu:2023ijx}), which predict the state of such matter and would constrain the high-energy limit of nuclear physics if confirmed or rejected.

It appears that the inspiral and merger of BNS systems offer a viable opportunity to uncover the precise physics of matter at this most extreme state~\cite{Flanagan:2007ix, Hinderer:2009ca, LIGOScientific:2017vwq, LIGOScientific:2018hze, LIGOScientific:2020aai}, in particular in tandem with future observations of, e.g., core-collapse supernovae~\cite{Morozova_2018}, NSBH systems~\cite{Abbott_2021}, and individual pulsars~\cite{Choudhury_2024}, to list a few. In BNS coalescence, the interior structure of NSs, particular its EoS, dictates how the NS reacts to a time-varying tidal field of a compact binary partner. Consequently, one can define the NS's EoS-dependent tidal deformability $\Lambda_l$, which quantifies the star's $l$-th order multipolar deformation $\mathcal{Q}$ in response to an external tidal field $\mathcal{E}$ as, e.g., $\mathcal{Q}_{ij}=\Lambda_2\mathcal{E}_{ij}$ for the quadrupole $(l=2)$~\cite{Flanagan:2007ix, Hinderer:2007mb}. An important goal of BNS detection is then to determine this deformability as well as the individual star's mass, spin, and potentially other NS quantities, to measure the correct EoSs.

Elaborating on one of these NS parameters, it has also been predicted that during the inspiral, the frequency of the time-varying tidal field can approach or cross the fundamental $f$-mode frequency of a NS. This can result in resonance phenomena that further augment the star's tidal response. These frequency-dependent \emph{dynamical tides} have been studied in Refs.~\cite{1994PThPh..91..871S, Flanagan:2007ix, Hinderer:2016eia, Steinhoff:2016rfi,Schmidt:2019wrl, Andersson:2019dwg, Gupta:2020lnv, Steinhoff:2021dsn, Pratten:2021pro, Kuan:2022etu, Pnigouras:2022zpx, Mandal:2023lgy, Mandal:2023hqa} and can be effectively described through separation-dependent tidal enhancement $\Lambda_l(r)=\keff(r)\Lambda_l$ and depend through the resonance frequency $\omega_{f,l}$ on the EoS as well as the NS's spin. In addition to the change of the $f$-mode frequency and dynamical response in the inertial frame of the star, the spin also induces a multipolar structure in BHs and NSs, which can be captured by higher-order spin effects~\cite{Henry:2022dzx,Khalilv5}, and affects the conservative and dissipative dynamics of BNS systems.

It is possible to analytically solve the approximate Einstein field equation of two point particles with some associated multipolar structure to include the aforementioned matter effects. In post-Newtonian (PN) theory (see, e.g., Refs.~\cite{Futamase:2007zz,Blanchet:2013haa,Schafer:2018kuf,Levi:2015msa,Porto:2016pyg,Levi:2018nxp,Isoyama:2020lls}
for reviews), Einstein's equations are expanded in powers of $(v/c)^{2}$, where $v$ is the binary characteristic velocity and $c$ is the speed of light. Within this framework, one can derive the effects of tidal interactions which start to affect the GW phase of a coalescing binary at the 5th PN order beyond the leading term ($\propto ({{v}/{c}})^{10}$)~\cite{Vines:2011ud} and are currently known beyond the next-to-next-to leading order or equivalently up to 7.5PN order~\cite{Damour:2012yf, Henry:2020ski, Narikawa:2023deu,Mandal:2024,Dones:2024}. The perturbative series resulting from these calculations has been shown to be accurate during the early inspiral phase~\cite{Messina:2019}. However, as the merger phase is approached, the series becomes inaccurate due to the significant relativistic velocities of the objects ($v\sim c$), the nonlinear deformation of NSs, and the potential for matter exchange~\cite{Bernuzzi:2016pie,Dietrich:2021}.

Consequently, numerical-relativity (NR) simulations have emerged as a valuable tool for modeling diverse matter phenomena when analytic methods become impractical, particularly in the vicinity of, and following the merger. NR predicts the GW emission of BNS systems by numerically solving the Einstein equations augmented with prescriptions for hydrodynamic and particle physics effects on supercomputers. However, as these simulations are computationally intensive, they generally encompass only the late stage of the inspiral and merger~\cite{Kawaguchi:2018gvj, Kiuchi:2019kzt, Foucart:2018lhe, Dietrich:2018phi, Ujevic:2022qle, Gonzalez:2022mgo}. 

Although these simulations are therefore our best effort at modeling the late stages of the coalescence of BNS systems, they are still prone to intrinsic errors, which must be paid careful attention to and which partially remain as open research issues. One typical error source for finite-difference codes are resolution effects, stemming from the numerical discretization of the Einstein equations and associated spacetime. Other error sources are, for example, the different phenomenological prescriptions to treat the hydrodynamics on these discretized grids; the omission of some physical effects like magnetic fields and their back reaction on the matter, which is the case for older simulations like the ones used in this work~\cite{Kiuchi:2022,Neuweiler:2024}; or the numerical heating of the NSs, which can distinguish numerical NSs from astrophysical ones~\cite{Fields:2023,Gittins:2025}.

A successful approach for combining PN results, and other analytical approximation methods, with information from high-accuracy NR simulations and strong-gravity effects has been the effective-one-body (EOB) formalism, see Refs.~\cite{Buonanno:1998gg,Buonanno:2000ef,Buonanno:2006ui,Damour:2000we,Damour:2001tu,Buonanno:2005xu}. It maps the dynamics of a compact binary to that of
a test mass (or test spin) in a deformed Schwarzschild (or Kerr) 
background, with the deformation parameter being the symmetric mass ratio $\nu$, supplemented with an associated energy loss due to GW emission. The resulting dissipative system of ordinary differential equations (ODE) is then integrated numerically.
EOB waveform models of BBHs have been constructed for non-spinning~\cite{Buonanno:2006ui,Buonanno:2007pf,Damour:2007yf,Damour:2008gu,Buonanno:2009qa,
  Pan:2011gk,Damour:2012ky,Damour:2015isa,Nagar:2019wds},
spinning~\cite{Damour:2001tu,Buonanno:2005xu,Damour:2007vq,
  Damour:2008qf,Pan:2009wj,Damour:2008te,Barausse:2009xi,Barausse:2011ys,Nagar:2011fx,Damour:2014sva,Balmelli:2015zsa,Khalil:2020mmr,Taracchini:2012ig,Taracchini:2013rva,Bohe:2016gbl,Cotesta:2018fcv,Pan:2013rra,Babak:2016tgq,Ossokine:2020kjp,Nagar:2018plt,Nagar:2018zoe,Akcay:2020qrj,Gamba:2021ydi,Pompiliv5}, and eccentric
binaries~\cite{Bini:2012ji,Hinderer:2017jcs,Chiaramello:2020ehz,Nagar:2021gss,Khalil:2021txt,Ramos-Buades:2021adz,Albanesi:2022xge,Gamboa:2024a}, and have recently been extended to scattering and dynamical captures~\cite{Nagar:2021,Andrade:2023,Albanesi:2025b,Albanesi:2025}. They are developed in two flavors, namely \texttt{SEOBNR}\ in Refs.~\cite{Bohe:2016gbl,Cotesta:2018fcv,Ossokine:2020kjp,Cotesta:2020qhw,Ramos-Buades:2021adz,Mihaylov:2021bpf}
and \teob\ in Refs.~\cite{Nagar:2018zoe,Nagar:2019wds,Nagar:2020pcj,Gamba:2021ydi,Riemenschneider:2021ppj,Chiaramello:2020ehz}. These models have also been extended to BNS systems, as \vfourt\ in Refs.~\cite{Hinderer:2016eia,Steinhoff:2016rfi} for the former and in Refs.~\cite{Bernuzzi:2014owa,Nagar:2018zoe,Akcay:2018yyh,Gonzalez:2022prs,Gamba:2023mww} for the latter. \teob\ is furthermore developed as \teobg\ for quasi-circular systems, and as \texttt{TEOBResumS-Dal\'i} for eccentric systems. For simplicity, and as we only study quasi-circular systems in this work, we refer to \teobg\ as the shorter \teob\ in the following.

As BNS waveforms are however in the sensitive frequency band of the detectors for thousands of wave-cycles, the pure ODE integration and subsequent Fourier transformation of the waveform can become too computationally intensive for PE. To overcome this issue, there exist reduced-order models of EOB approximants, see Refs.~\cite{Lackey:2016krb,Lackey:2018zvw}, or one can use the post-adiabatic approximations as in Refs.~\cite{Nagar:2018gnk,Mihaylov:2021bpf}, machine learning techniques as in Refs.~\cite{ Tissino:2022thn}, and the stationary phase approximation instead of a Fast Fourier Transform of the waveform, see Ref.~\cite{Gamba:2020ljo}. 

There also exist phenomenological GW models in Refs.~\cite{Ajith:2009bn, Santamaria:2010yb}, which combine and extrapolate PN, EOB and NR waveforms across the parameter space of interest. The most important phenomenological BNS waveform approximant is the \nrtidal\ family, developed in Refs.~\cite{Dietrich:2017aum, Dietrich:2019kaq,Kawaguchi:2018gvj,Abac:NRT}. It aims to model the tidal phase contribution during the inspiral through the use of closed-form analytical expressions for the tidal contribution and an NR calibrated merger model. The matter contribution is then used to augment a given BBH waveform baseline to model BNS systems.

In this work, we introduce \vfivet\ 
as the next step in modeling spin-aligned BNS systems accurately and fast in the \texttt{SEOBNR} framework. We have improved the model in comparison to \vfourt\ through
\begin{enumerate}[label=(\roman*)]
\itemsep-3pt
\item the development of \vfivet\ in the highly optimized \texttt{pySEOBNR} python package and with the modeling improvements of the point-particle case as developed in \vfivehm~\cite{Pompiliv5}, which yields more accurate BNS waveforms that are faster by a factor 100 to 1000 when comparing to \vfourt\ for systems with total mass $M \geq 2\, M_\odot$\footnote{Note that this applies to the comparison between the EOB models \vfourt\ and \vfivet. There also exists the frequency-domain surrogate \texttt{SEOBNRv4T\_surrogate}~\cite{Lackey:2018zvw} for applications in PE, which is even faster than \vfivet, as discussed in Sec.~\ref{subsec:PA}.};
\item analytical improvements in the inclusion of up to relative 2.5PN order adiabatic tidal effects~\cite{Henry:2020ski,Dones:2024,Mandal:2024}, the inclusion of 3.5PN and 4PN order spin-induced multipolar effects for spin-squared, spin-cubed and spin-to-the-fourth contributions in the conservative sector~\cite{Khalilv5} and in the dissipative sector~\cite{Henry:2022dzx}, as well as dynamical tides for spinning NSs~\cite{Flanagan:2007ix, Hinderer:2016eia, Steinhoff:2016rfi, Schmidt:2019wrl,Steinhoff:2021dsn};
\item development of an NR informed phenomenological pre-merger model; and in particular
\item the calibration to a large set of NR BNS waveforms in the merger time. 
\end{enumerate}

In total, we utilize 64 NR waveforms from the BAM catalogues~\cite{Dietrich:2018phi,Gonzalez:2022mgo, Ujevic:2022qle} and the SACRA catalogues~\cite{Kiuchi:2017pte, Kawaguchi:2018gvj}, covering 10 different EoSs, with mass ratios $q\in [1,\,2]$, tidal deformabilities $\Lambda\in [43,\,4090]$, and dimensionless spins $\chi\in [-0.18,\,0.27]$.

The paper is organized as follows. In Sec.~\ref{sec:EOB}, we summarize the EOB framework and present how we include tidal effects in the Hamiltonian, in the radiation-reaction-force, and the waveform of \vfivehm\ to accurately model BNS systems and explain our pre-merger modeling approach. We also discuss the applicability of our model to highly spinning, anti-aligned NSs, which can exhibit $f$-mode resonances already during the early inspiral. In Sec.~\ref{sec:Calibration} we lay down our approach for the calibration against NR waveforms, where we pay close attention to the resolution of different NR runs and extrapolate our results to a common resolution across all employed simulations. In Sec.~\ref{sec:Validation} we discuss our implementation and perform tests to validate our model through comparisons to NR, and to the other state-of-the-art waveform models across the relevant parameter space for observations. We further perform PE for the two BNS detections GW170817 and GW190425 in Sec.~\ref{sec:PE} and compare our results to the literature.
Finally, our main conclusion, directions for future improvements of the model and recommendations for the region of validity of \vfivet\ are discussed in Sec.~\ref{sec:Conclusion}.

\section*{Notation}
In the following, we use natural units in which $c = G =1$ unless stated otherwise, and we use overhead dots to signal time derivatives. We largely follow the notation as outlined in Ref.~\cite{Pompiliv5} and consider binaries with masses $m_1$ and $m_2$ where $m_1 \geq m_2$.
We also define the following derived quantities:
\begin{equation}
\begin{gathered}
	M\equiv m_1 + m_2, \qquad q \equiv \frac{m_1}{m_2} \geq 1, \qquad \nu \equiv \frac{m_1 m_2}{M^2},\\
	\delta \equiv\frac{m_1 - m_2}{M},  \qquad \mu \equiv \frac{m_1m_2}{M}, \qquad \mathcal{M} = \nu^{3/5}M \\ 
    \quad X_i \equiv \frac{m_i}{M}, \qquad X=1-4\,\nu.
\end{gathered}
\end{equation}

For NSs we define the adiabatic tidal deformability $\Lambda_l^{(i)}$ in accordance with the masses $i\in\{1,2\}$. It is related to the dimensionless tidal Love number $k_l^{(i)}$ and the compactness $\mathcal{C}_{i}=m_i/R_i$ defined through the NS radius $R_i$ as
\begin{subequations}
    \begin{align}
        \Lambda_l^{(i)} &= \frac{2}{(2l-1)!!}\frac{k_l^{(i)}}{\mathcal{C}_{i}^{2l+1}} \label{eq:lambda_m}\ , \\
        \qquad &= \frac{2}{(2l-1)!!}k_l^{(i)}\frac{(R_i/M)^{2l+1}}{X_i^{2l+1}}\ . \label{eq:lambda_M}
    \end{align}
\end{subequations}
It should be highlighted that the tidal deformability as normalized through the compactness $\mathcal{C}_i$ in Eq.~\eqref{eq:lambda_m} is given in terms of the individual masses $m_i$. If we want to work in units of total mass~$M$, we therefore need to use $\Lambda_l^{(i)}\, X_i^{2l+1}$ as the dimensionless adiabatic tidal deformability in units of total mass, as can be seen from Eq.~\eqref{eq:lambda_M}.

We also define the quadrupolar tidal coupling constant
\begin{equation}
    \begin{gathered}
        \kappa_2^T = 3\nu \left(X_1^3 \Lambda_2^{(1)}
        + X_2^3 \Lambda_2^{(2)}\right),
    \end{gathered}
    \end{equation}
which quantifies the leading order contribution of tidal effects.

For binaries with non-precessing spins of magnitude $S_1$ and $S_2$, we define the dimensionless spins
\begin{gather}
	\chi_{ i} \equiv \frac{a_{ i}}{m_{ i}} = \frac{S_{ i}}{m_{ i}^2} \in [-1,\,1],
\end{gather}
where $ i \in \{1,2\}$, and define the following spin variables:
\begin{equation}
\begin{gathered}
    \label{eq:spins}
	\chi_{\mathrm{eff}} \equiv \frac{\left(m_1 \chi_1+m_2 \chi_2\right)}{m_1+m_2}, \\
	a_\pm \equiv a_1 \pm a_2 = m_1 \chi_1 \pm m_2 \chi_2.	
\end{gathered}
\end{equation}

We denote the spin-induced quadrupole, octupole, and hexadecupole constants as $C_2$, $C_3$, and $C_4$ respectively. These constants equal 1 for BHs but are larger than 1 for NSs. We occasionally use $u \equiv M/r$ instead of the separation $r$, and introduce the Keplerian orbital frequency $\Omega_0 \equiv (r/M)^{-3/2}$.

Our conventions for the Fresnel integrals are
\begin{equation}
    \label{eq:Fresnel}
    \begin{split}
        F_S(z) &= \int_0^z \sin \left(\frac{\pi t^2}{2}\right) \d t, \\ F_C(z) &= \int_0^z \cos \left(\frac{\pi t^2}{2}\right) \d t.
    \end{split}
\end{equation}

As in the BBH case, we define the \emph{merger time} of an aligned-spin BNS waveform to be the time where the $(2,2)$ mode amplitude first peaks.

\section{The SEOBNRv5THM model} \label{sec:EOB}

In the following section, we describe our EOB model, highlighting the changes we make to the BBH baseline \vfivehm\ from Ref.~\cite{Pompiliv5}. More specifically, in Sec.~\ref{subsec:Hamiltonian} we explain how we include spin-induced multipole moments and dynamical tides into the Hamiltonian of our model, and the changes we make to the dynamical tide computation in comparison to \vfourt. In Sec.~\ref{subsec:Waveform} we then describe how the inspiral waveform is computed, and how we include relative 2.5PN adiabatic tidal and 3.5PN spin-induced multipole information into the inspiral waveform and fluxes. In Sec.~\ref{subsec:Merger} we explain our novel approach to the termination and pre-merger waveform for BNS systems.

\subsection{EOB dynamics with matter effects} \label{subsec:Hamiltonian}

The EOB formalism~\cite{Buonanno:1998gg,Buonanno:2000ef,Damour:2000we,Damour:2001tu,Buonanno:2005xu} maps the two-body dynamics onto the effective dynamics of a test body in a deformed Schwarzschild or Kerr background, with the deformation parametrized by the symmetric mass-ratio $\nu$.
The energy map relating the effective Hamiltonian $H_{\rm{eff}}$ and the two-body EOB Hamiltonian $H_{\rm{EOB}}$ is given by
\begin{equation}
\label{eq:HEOB}
	H_{\mathrm{EOB}}=M \sqrt{1+2 \nu \left(\frac{H_{\mathrm{eff}}}{\mu}-1\right)}\,.
\end{equation}

We use the aligned-spin \vfivehm\ waveform model as our BBH baseline and refer interested readers to Ref.~\cite{Pompiliv5} for a more in-depth description of the model. In the following, we summarize the structure of the aligned-spin Hamiltonian, and its zero-spin limit, highlighting where matter effects enter into the model.

For aligned spins, the effective Hamiltonian reduces to the equatorial Kerr Hamiltonian in the test-body and BBH limit, with the Kerr spin $a=a_+$ mapped to the binary's total spin. It explicitly reads~\cite{Pompiliv5,Khalilv5}
\begin{align}
\label{HeffAnzAlign}
&H_\text{eff}^\text{align} \nonumber \\
&\equiv \frac{Mp_\phi \left(g_{a_+} a_+ + g_{a_-} \delta a_-\right) + \text{SO}_\text{calib}
+ G_{a^3}^\text{align}}{r^3+a_+^2 (r+2M)} \\
&
+  \sqrt{A^\text{align} \Bigg(
\mu^2 + p^2 + B_{np} p_r^2 + B_{npa} \frac{p_\phi^2 a_+^2}{r^2} + Q^\text{align}
\Bigg) }, \nonumber
\end{align}
where $r$ is the binary's separation and $\phi$ the orbital phase, $p_r$, $p_\phi$ are their canonically conjugated momenta that satisfy $p^2=p_r^2+p_\phi^2/r^2$, $g_{a_\pm}$ are gyro-gravitomagnetic factors, $A^\text{align}$, $B_{np(a)}$, $G_{a^3}^\text{align}$, and $Q^\text{align}$ are EOB potentials containing resummed PN information, defined in Appendix A of Ref.~\cite{Pompiliv5}, and $\text{SO}_\text{calib}$ is a calibration parameter tuned to NR. The first term on the right-hand side includes the odd-in-spin contributions (in the numerator), while the second term (square root) includes the even-in-spin contributions. Some parts of the matter information influence the $A^\text{align}$-potential, which reads more explicitly
\begin{equation}
    A^\text{align} = \frac{a_+^2/r^2+A_\text{noS}+A_\text{SS}^\text{align}}{1+(1+2M/r)a_+^2/r^2},
\end{equation}
where $A_\text{noS}$ and $A_\text{SS}^\text{align}$ are further potentials that contain PN information about the non-spinning and spin-spin (SS) sector.

We remind that such a Hamiltonian includes most of the 5PN non-spinning contributions, together with spin-orbit
(SO) information up to the next-to-next-to-leading order (NNLO), SS information to NNLO,
as well as cubic- and quartic-in-spin terms at LO, corresponding to all PN information up to 4PN order for spin aligned point particles. More details about the derivation of the generic-spin Hamiltonian, together with the full expressions of all terms and potentials, are given in Ref.~\cite{Khalilv5}.

Regarding matter effects, the \vfivet\ Hamiltonian includes the spin-induced multipole moments $C_l$ up to $l=4$, to the same order as the BBH contributions, which in Ref.~\cite{Khalilv5} are defined as $C_{\rm iES2},\ C_{\rm iBS3},$ and $C_{\rm iES4}$ respectively. These terms have been set to unity for the BBH baseline model \vfivehm, where spin-induced multipoles are absent, but enter now in the $A^{\rm align},\ B_{np},$ and $Q^{\rm align}$ expressions for $l=2,4$ and the $G^{\rm align}_{a^3}$ expression for $l=3$ as given explicitly in Eqs. (31)--(34) of Ref.~\cite{Khalilv5}.

In the non-spinning (noS) limit $a_1=a_2=0$, the effective Hamiltonian reduces to~\cite{Khalilv5}
\begin{equation}
    \label{Heffzero}
    H_\text{eff}^\text{noS} = \sqrt{A_\text{noS}\left[\mu^2 + A_\text{noS} \bar{D}_\text{noS}\, p_r^2 + \frac{p_\phi^2}{r^2} + Q_\text{noS}\right]}\,,
\end{equation}
and the 5PN potentials $A_\text{noS}(r)$ and $\bar{D}_\text{noS}(r)$ are complete except for two quadratic-in-$\nu$ coefficients. They are respectively (1,5) and (2,3) Pad\'e resummed. $A_\text{noS}$, $\bar{D}_\text{noS}(r)$ and $Q_\text{noS}$ are furthermore respectively contained in $A^{\rm align}$, $B_{np}$, and $Q^{\rm align}$ from Eq.~\eqref{HeffAnzAlign}.

As has been done for \vfourt\ in Ref.~\cite{Lackey:2018zvw}, we add NNLO dynamical tidal effects for the quadrupole ($l=2$) and octupole ($l=3$) into the Hamiltonian through the modification of the BBH $A_\text{noS}$-potential, see Ref.~\cite{Bini:2012a}
    \begin{equation}
    \label{eq:AnoS}
    A_\text{noS}(r) \mapsto A_\text{noS}^{\rm BBH}(r) + A_\text{noS}^{\rm tidal}(r), 
\end{equation}
where $A_\text{noS}^{\rm BBH}(r) \equiv A_\text{noS}(r)$ of the BBH baseline and 
\begin{widetext}
    \begin{equation}
    \begin{split} \label{eq:Atidal}
     A_\text{noS}^{\rm tidal}(r) =&  - 3\,\Lambda_2^{(1)} k^\mathrm{eff, (1)}_{2}(r) X_1^4 X_2\,u^6  \left[ 1 + \frac{5}{2} X_1 u + \left(3 + \frac{1}{8}X_1 + \frac{337}{28} X_1^2\right)u^2 \right] \\
     &- 15\,\Lambda_3^{(1)} k^\mathrm{eff, (1)}_{3}(r) X_1^6X_2\,u^8 \left[1 + \left(-2 + \frac{15}{2}X_1\right)u + \left(\frac{8}{3} - \frac{311}{24}X_1 + \frac{110}{3}X_1^2\right) u^2\right] + (1 \leftrightarrow 2)\,.
    \end{split}
    \end{equation}
\end{widetext}
In the above expression, $k^{\mathrm{eff}, (i)}_{l}(r)$ are the effective tidal enhancement factors due to dynamical tides, defined below in Eq.~\eqref{eq:keff}.

Note that we are only including relative 2PN adiabatic tidal information (up to $u^2$) in the Hamiltonian in Eq.~\eqref{eq:Atidal}. Our effective description of dynamical tides in $k^{\mathrm{eff}, (i)}_{l}(r)$ on the other hand is based on a leading-order two-timescale approximation~\cite{Steinhoff:2016rfi,Steinhoff:2021dsn}. Currently, the relative 3PN dynamical and adiabatic tidal correction to the Hamiltonian is known~\cite{Mandal:2023hqa,Mandal:2023lgy}, but the calculation requires a counterterm whose finite contribution has yet to be determined for specific EoSs, and the Hamiltonian is not yet EOB resummed. These results can therefore not be immediately used in EOB models.

We note that although one could, as is done in \teob\ in Refs.~\cite{Gonzalez:2022prs,Gamba:2023mww}, include the LO adiabatic tidal effects for $l\geq 4$, gravitomagnetic tidal effects, and the respective tidal adjustment of the $\bar{D}_\text{noS}$-potential, we leave the inclusion of these effects as future work. Our reasoning for this is that those effects were partially known but not included in \vfourt\ either, are outside the scope of this work, and have little impact on the waveform as one can see later in Sec.~\ref{sec:Validation}. We furthermore believe it to be more consistent to include only adiabatic information where the dynamical tidal extension is currently understood, which is not the case for the aforementioned adiabatic tidal information.

Our computation of the effective tidal enhancement factor $k^\mathrm{eff}_l(r)$ to be used in the $A_\text{noS}^{\rm tidal}$ potential of Eq.~\eqref{eq:Atidal} for spinning NSs is based on a two-timescale expansion between the orbital and oscillation timescales and taken from Ref.~\cite{Steinhoff:2021dsn}. We however rewrite it here in a more condensed form inspired by the non-spinning $k^\mathrm{eff}_l(r)$ as reported in Ref.~\cite{Lackey:2018zvw}. It reads
\begin{equation}
    \label{eq:keff}
    k^\mathrm{eff}_l(r) = a_l + b_l \left[f(x_f,\Omega_0)+\left(1-\frac{\Delta\omega_{0l}^2}{\omega_{0l}^2}\right)\sqrt{\frac{\pi}{3\varepsilon}}Q(\hat{t})\right],
\end{equation}
where the coefficients for the quadru- and octupole multipole moments are $(a_2,\, a_3) = (1/4,\, 3/8),$ and $(b_2,\, b_3) = (3/4,\, 5/8)$, the resonant term $f$ and the trigonometric term $Q$ are defined in Eqs.~\eqref{eq:f}--\eqref{eq:keff_defs} below. We additionally omit the index $(i)$ labeling the NS, but note that $\keff$ has to be computed for each of them individually before it can be used in Eq.~\eqref{eq:Atidal}.

As we employ the Keplerian orbital frequency $\Omega_0(r)$, it ultimately only depends on the separation $r$ in addition to the NS parameters $(\Lambda_{l \geq 2}, \omega_{0l}, \Delta \omega_{0l})$, where $\omega_{0l}$ is the $f$-mode frequency of the $2^l$ multipole for a non-spinning NS, and $\Delta \omega_{0l}$ is the shift in $f$-mode frequency due to the NS's spin. We denote the resulting $f$-mode frequency for spinning NSs by $\omega_{f,l}$, see Eq.~\eqref{eq:keff_defs} below. All frequencies are given in units of inverse total mass $M^{-1}$ in the equations below.
The remaining equations for the computation of $\keff (r)$ then read
\begin{widetext}
    \begin{subequations}
        \begin{equation} \label{eq:f}
            f(x_f,\Omega_0) = \frac{(1-\frac{\Delta\omega_{0l}^2}{\omega_{0l}^2})}{1 - \frac{(l  \Omega_0 -\Delta\omega_{0l} )^2}{\omega_{0l}^2}} + \frac{5}{6}\left(1-\frac{\Delta\omega_{0l}}{\omega_{0l}}\right)\frac{x_f^2}{(1-x_f^{5/3})},
        \end{equation}
    \begin{equation}
            Q(\hat{t}) = \cos\left(\frac{3}{8}\hat{t}^2\right) \left[1+2F_S\left(\frac{\sqrt{3}}{2\sqrt{\pi}}\hat{t}\right)\right] -\sin\left(\frac{3}{8}\hat{t}^2\right) \left[1+2F_C\left(\frac{\sqrt{3}}{2\sqrt{\pi}}\hat{t}\right)\right],
        \label{eq:Q}
    \end{equation}
    \begin{equation}
        \begin{gathered} \label{eq:keff_defs}
            \omega_{f,l}=\omega_{0l} + \Delta\omega_{0l}, \quad x_f=\frac{\omega_{f,l}}{l\Omega_0}, \quad \epsilon=\frac{256\nu}{5}\left(\frac{M\omega_{f,l}}{l}\right)^{5/3}, \quad \hat{t}=\frac{8}{5\sqrt{\varepsilon}}\left[1-x_f^{5/3}\right],
        \end{gathered}
    \end{equation}
    \end{subequations}
\end{widetext}
where $x_f$ is the ratio between the forced $f$-mode and driving orbital frequency, $\epsilon$ is a small dimensionless parameter used in the two-timescale expansion~\cite{Steinhoff:2021dsn} and $\hat{t}\equiv\sqrt{\epsilon}(\phi - \phi_{\mathrm{res}})$ is a rescaled shifted dimensionless phase variable that describes the orbital phasing around resonance, denoted by `res'. We remind that the Fresnel integrals used in Eq.~\eqref{eq:Q} are defined in Eq.~\eqref{eq:Fresnel}.

It has to be noted that the $f(x_f)$ of Eq.~\eqref{eq:f} has a removable singularity for the resonance $x_f=1 \iff l\Omega_0 = \omega_{f,l}$. Similarly to what has been done in \vfourt, we take care of this singularity by switching to the fourth order Taylor expansion of $f$ in the vicinity of the pole $|x_f-1|<10^{-2}$, while keeping $\omega_{0l}$ and $\Delta \omega_{0l}$ fixed and replacing $\Omega_0 = \omega_{f,l}/(lx_f)$ prior to the expansion. This procedure only introduces a negligible discontinuity at the transition region of the order $10^{-10}$.

By close inspection of Eq.~\eqref{eq:keff_defs}, we find that 
\begin{equation}
    \hat{t}\propto -(r/M)^{5/2} + \rm{const.}
\end{equation}
grows exponentially for large orbital separations $r$. As a consequence, the trigonometric functions and Fresnel integrals used in the definition of $Q(\hat{t})$ of Eq.~\eqref{eq:Q} become numerically unstable for large separations $r$ during the early inspiral in both the spinning and non-spinning case.

In contrast to what has been done in \vfourt, we therefore use the Taylor expansion of $Q$ around $1/\hat{t}=0$, or equivalently $r\rightarrow +\infty$
\begin{equation} 
        \label{eq:1}
        Q(\hat{t}) = -\frac{4}{\sqrt{3\pi }}\frac{1}{\hat{t}} + \mathcal{O}(\hat{t}^{-5}),
\end{equation}
for the early inspiral, where the tides behave nearly adiabatically. In the interval $\hat{t}\in [-45,-44]$ we then transition to the full Eq.~\eqref{eq:Q} for $Q(\hat{t})$. We have decided on this transition interval for $\hat{t}$ as through this approach all relative and absolute errors of $Q(\hat{t})$ and $Q'(\hat{t})$ are below $10^{-4}$.

We want to highlight that this change alone increases the waveform generation speed for long waveforms starting at 20 Hz by close to two orders of magnitude compared to \vfourt, as the adaptive ODE solver is not picking up numerical noise. It should additionally increase the model's accuracy in the early inspiral.

We further employ quasi-universal relations as given in Refs.~\cite{Yagi:2014bxa,Yagi:2016bkt,Kruger:2020,Sotani:2021,Gamba:2021prd,Kruger:2023} to relate the free NS parameters $(\Lambda_{l > 2}, \omega_{0l}, \Delta \omega_{0l}, C_l)$ in the model to the tidal deformability $\Lambda_2$. This way, we can compute accurate waveforms for different EoSs. One can however also specify those parameters freely for uses in PE or to study their impact on the waveform. A more precise description of the quasi-universal relations used for \vfivet\ can be found in Appendix~\ref{app:URs}.

It is known that for anti-aligned (counter-rotating) NSs the total $f$-mode frequency $\omega_{f,l}$ can become zero and even negative for high spins, which makes the $f$-mode unstable.  In particular, Eq.~\eqref{eq:f} can develop a proper singularity in the denominator of the first term for $l  \Omega_0 -\Delta\omega_{0l} = \omega_{0l}$ for large, negative $\Delta\omega_{0l}$ associated to counter-rotating NSs. Such a singularity makes the ODE integration impossible, but has not been considered previously for \vfourt. This regime of large (anti-aligned) spins requires an extension of the model, possibly through the addition of nonlinear dynamical tidal effects, which we leave for future investigations. 

This effect is essentially the handrasekhar-Friedman-Schutz, CFS instability (see Refs.~\cite{CFS:1,CFS:2}), where counter-rotating linear perturbations on a spinning NS can become unstable to the emission of GWs. This leads to a spin-down of the NS over secular timescales, much shorter than the inspiral timescale, until $\omega_{f,l} > 0$ is reached and the instability disappears. This effect, in tandem with other instabilities like the $r$-mode instability in nascent stars, therefore limits the maximum allowed spin-rate of astrophysical NSs expected in BNS systems, see the extensive reviews of Refs.~\cite{Andersson:2003,Paschalidis:2016vmz}. 

In order to still be able to generate waveforms across the full parameter space for this work, we apply a smooth asymptotic cutoff on the spin-shift to ensure $|\Delta \omega_{0l} / \omega_{0l}| < 0.9$ on the $\Delta \omega_{0l}$ as taken from Ref.~\cite{Kruger:2020}.
While this generated waveform in the instability regime is unphysical, this regime is also astrophysically implausible for NSs in inspiraling binaries. The observed maximum NS spin rates furthermore suggest the validity of our procedure~\cite{Hessels:06,Nielsen:2016}, given that they hint at a maximum spin rate below the onset of the CFS instability for the $l\in\{2,\,3\}$ $f$-modes~\cite{Paschalidis:2016vmz}.

Recently, the authors of Ref.~\cite{Yu:2024} have studied anti-aligned systems for the dynamical tides in the formulation of Ref.~\cite{Steinhoff:2021dsn} for highly spinning anti-aligned NSs and followed up on their work in Ref.~\cite{Yu:2025}. They find that the crossing of the resonance can induce eccentricity in the originally quasi-circular system. We find the same development of eccentricity when naively continuing the integration of the EOB system of ODEs. The resulting waveform of such a continued integration is however clearly unphysical due to the assumption of quasi-circularity in the model and in particular in the quasi-circular derivation of $\keff$ from Refs.~\cite{Steinhoff:2016rfi,Steinhoff:2021dsn}. It was also shown before in Ref.~\cite{Gamba:2022mgx} that the current description of dynamical tides in terms of the effective (circular-orbit) tidal deformability $\keff$ is not accurate in modeling eccentric BNS systems. While generic-orbit EOB Hamiltonians not based on $\keff$ have been derived~\cite{Steinhoff:2016rfi}, a complete dynamical-tidal EOB waveform model for eccentric orbits in missing. 

As our model assumes quasi-circularity and because we do not introduce the tidal torque that is suggested in Ref.~\cite{Yu:2024} or the tidal spin that is discussed in Ref.~\cite{Yu:2025}, we have opted to terminate the waveform prior to the development of eccentricity, as soon as $\dot{p}_{r_*} > 0$.

In contrast, we want to note that the authors of Ref.~\cite{Kuan:2024} have NR simulated one such BNS system with large anti-aligned spins with the SACRA code and found no signs of eccentricity developing. Instead, they report a change of the spins of the individual NSs as they go through resonance and gain spin angular momentum from the orbit due to tidal torques. Currently, there only exists the very recent work of Ref.~\cite{Yu:2025} investigating the magnitude of the dynamical tidal contribution to the spin, which could lead to angular momentum transfer during $f$-mode resonance, but has not been verified against NR.  We therefore note the open question of the waveform of high spinning NSs and leave the development and inclusion of a consistent framework for angular momentum exchange due to $f$-mode resonances as future work.

For now, we suggest to only use the model up to medium NS spins of $|\chi_{\rm NS}| \lesssim 0.3$ for this reason.

The resulting Hamiltonian obtained after incorporating all relevant matter information can then be used to solve the equations of motion of the aligned-spin BNS system. They are given in Refs.~\cite{Pompiliv5,Pan:2011gk} and read
\begin{subequations}
    \begin{equation}
    \label{eq:eom_prst}
    \dot{r} = \xi \frac{\partial H_{\rm{EOB}}}{\partial p_{r_*}}, \quad
    \dot{p}_{r_*} = -\xi \frac{\partial H_{\rm{EOB}}}{\partial r} + \frac{p_{r_*}}{p_\phi} \mF_\phi,
    \end{equation}
    \begin{equation}
    \label{eq:eom_pphi}
    \dot{\phi} = \frac{\partial H_{\rm{EOB}}}{\partial p_\phi}, \qquad
    \dot{p}_\phi = \mF_\phi,
    \end{equation}
\end{subequations}
where $p_{r_*}$ and $\xi$ relate to the tortoise coordinates defined below, and the RR force $\mF_\phi$ is obtained by summing the GW modes in factorized form $h_{\ell m}^{\mathrm{F}}$ as in Refs.~\cite{Damour:2007xr,Damour:2007yf,Damour:2008gu,Pan:2010hz}, which we define in Sec.~\ref{subsec:Waveform}, that is
\begin{equation}
\label{eq:RRforce}
\mF_\phi \equiv - \frac{M \Omega}{8 \pi} \sum_{\ell=2}^8 \sum_{m=1}^{\ell} m^2\left|d_L h_{\ell m}^{\mathrm{F}}\right|^2,
\end{equation}
where $\Omega = \dot{\phi}$ is the orbital frequency, and $d_L$ is the luminosity distance of the binary to the observer.

Like in the BBH baseline, we use the tortoise-coordinate $p_{r_*}$ instead of $p_r$,
since it improves the stability of the equations of motion during the 
plunge and close to merger~\cite{Damour:2007xr,Pan:2009wj,Pompiliv5}.
The tortoise-coordinate $r_*$ is defined through
\begin{equation}
    \frac{dr_*}{dr} = \frac{1}{\xi(r)}
\end{equation}
and its conjugate momentum is given by
\begin{equation}
    p_{r_*}=\xi(r)p_r.
\end{equation}
We further choose $\xi(r)$ as
\begin{equation} \label{eq:xi}
\xi(r) = \frac{\sqrt{\bar{D}_\text{noS}}\left(A_\text{noS} + a_+^2/r^2\right)}{1 + a_+^2/r^2},
\end{equation}
Note that $\xi$ has been constructed to reduce to the Kerr value $(dr/dr_*) = (r^2-2Mr+a_+^2)/(r^2+a_+^2)$ in the test particle limit $\nu\to 0$.

As in \vfivehm, we use quasi-circular adiabatic initial conditions derived in Ref.~\cite{Buonanno:2005xu}, and
then solve for the binary dynamics by numerically integrating Eqs.~\eqref{eq:eom_prst} and \eqref{eq:eom_pphi} using an $8^{\rm th}$ order Dormand-Prince integrator.

\subsection{EOB inspiral-waveforms with matter effects} \label{subsec:Waveform}

Like the Hamiltonian, our waveforms are built on the BBH \vfivehm\ multipolar waveform~\cite{Pompiliv5}. In the following, we give an overview of the waveform computation highlighting our changes to include matter effects.

In general, the complex linear combination of GW polarizations,  $h(t) \equiv h_{+}(t) -ih_{\times}(t)$, can be expanded in the basis 
of $-2$ spin-weighted spherical harmonics~\cite{Pan:2011gk} as follows:
\begin{equation}
	h(t;  \bm{\lambda}, \iota, \varphi_0) = \sum_{\ell \geq 2}\sum_{|m|\leq \ell} {}_{-2} Y_{\ell m} (\iota, \varphi_0) \, h_{\ell m}(t;\bm{\lambda}),
\label{eq:hoft_sphericalH}
\end{equation}
where $\bm{\lambda}$ denotes the intrinsic parameters of the compact binary, such as mass ratio $q$, spins $\chi_{1,2}$, and the NS parameters $(\Lambda_{l \geq 2}, \omega_l, \Delta \omega_l, C_l)$.
The parameters $(\iota, \varphi_0)$ describe the binary's inclination angle (computed with respect to the 
direction perpendicular to the orbital plane) and the azimuthal direction to the observer, respectively.

In \vfivet, the GW modes defined in Eq.~(\ref{eq:hoft_sphericalH})
are decomposed into inspiral modes and pre-merger modes close to merger. All modes are further tapered to zero after the merger. \vfivet\ can compute and return the waveforms of the dominant (2,2) mode and the largest subdominant modes: (3,3), (2,1), (4,4) (3,2), (5,5), and (4,3). 

In this work, we model the pre-merger of the dominant (2,2) mode phenomenologically up to merger, as we will describe in detail in Sec.~\ref{subsec:Merger}. For the subdominant modes however, our pre-merger waveform only consists of a tapering of the inspiral-waveform close to and after merger, which we discuss in Appendix~\ref{app:tapering}.

For aligned-spin binaries it holds that $h_{\ell m}=(-1)^{\ell} h_{\ell-m}^*$, therefore we restrict the discussion to $(\ell,m)$ modes with $m > 0$.
We have: 
\begin{equation}
	\label{eq:h_match}
	h_{\ell m}(t)= \begin{cases}h_{\ell m}^{\text {inspiral}}(t), & t<t_{\text {trans}} \\ h_{\ell m}^{\text {pre-merge}}(t), & t\geq t_{\text {trans }}\end{cases},
\end{equation}
where we define the transition time $t_{\text {trans}}$ and $h_{\ell m}^{\text {pre-merge}}(t)$ later in Sec.~\ref{subsec:Merger}.

The factorized inspiral modes are written as~\cite{Pompiliv5}
\begin{equation}
\label{eq:hlmFactorized}
	h_{\ell m}^{\mathrm{F}}=h_{\ell m}^{\text{N}} \hat{S}_{\mathrm{eff}} T_{\ell m} f_{\ell m} e^{i \delta_{t m}}.
\end{equation}
where details regarding the individual factors can be found in Eqs. (26)--(33) of Ref.~\cite{Pompiliv5}. The different factors in the modes depend on the set of ODE dynamical quantities ($r,\ \phi,\ \Omega$).

We include tidal information into the amplitude $f_{\ell m}$ and phase factors $\delta_{\ell m}$ of the factorized modes in Eq.~\eqref{eq:hlmFactorized}, which are computed such that the expansion of $h_{\ell m}^\text{F}$ agrees with the PN-expanded modes.
For non-spinning binaries, $f_{\ell m}$ is further resummed as $f_{\ell m} = (\rho_{\ell m})^\ell$ to reduce the magnitude of the 1PN coefficient, which grows linearly with $\ell$~\cite{Damour:2008gu}.
Following Refs.~\cite{Pan:2010hz,Taracchini:2012ig,Taracchini:2013rva}, we separate the non-spinning, spin and tidal contributions for the odd $m$ modes, such that
\begin{align}
\label{frholm}
f_{\ell m} = \left\{
        \begin{array}{ll}
           \left(\rho_{\ell m}^{\rm noS} + \rho_{\ell m}^{\rm S} + \rho_{\ell m}^{\rm tidal}\right)^\ell, & \quad m \text{ even}, \\\\
           (\rho_{\ell m}^\text{noS})^\ell + f_{\ell m}^\text{S} + f_{\ell m}^\text{tidal}, & \quad m \text{ odd},
        \end{array}
    \right. 
\end{align}
where $\rho_{\ell m}^\text{noS}$ ($\rho_{\ell m}^\text{S}$) is the non-spinning (spinning) part of $\rho_{\ell m}$, while $f_{\ell m}^\text{S}$ is the spin part of $f_{\ell m}$ and $f_{\ell m}^\text{tidal}$ as well as $\rho_{\ell m}^\text{tidal}$ contain the factorized form of the relative 2.5PN adiabatic gravitoelectric tidal modes from Ref.~\cite{Henry:2020ski}. The latter were found to be partially incorrect and were corrected in Refs.~\cite{Mandal:2024,Dones:2024}, and are given in their factorized form in Ref.~\cite{Dones:2024} and its supplementary material.

For the other tidal EOB models \teob\ and \vfourt, the tidal information in the modes has instead been added to the factorized point-mass modes through an ansatz of the form
\begin{equation}
    h_{\ell m} = h_{\ell m}^{\rm F} + h_{\ell m}^{\rm tidal} .
\end{equation}
We highlight that \vfivet\ is therefore currently the only EOB model where tidal effects are instead included in the modes in their factorized form. 

We employ the factorized $f_{\ell m}^\text{S}$ modes for the spin-induced multipole moments from Ref.~\cite{Henry:2022dzx}, Sec. IV.B, Appendix B, as well as its supplementary material. We include the spin-induced multipole moments in the modes consistently to their inclusion in the Hamiltonian up to 3.5PN order in SS, S$^3$ and S$^4$.

In comparison with \vfourt\ however, we have decided to omit the effective tidal enhancement of the adiabatic tidal modes due to dynamical effects from our model (\cite{Lackey:2018zvw} and Eq.~(16) of Ref.~\cite{Dietrich:2017}). This is because the expression used in \texttt{lalsimulation} is different from its published counterparts in Refs.~\cite{Dietrich:2017,Dietrich:2020efo,Gamba:2022mgx}. As this discrepancy on how the dynamical tides should affect the waveform's modes is not resolved, and as this effect only becomes relevant towards merger, which we model phenomenologically, we have decided to omit this for now.  We plan to revisit the effect of dynamical tides on the waveform modes in later work.

For the odd $m$ modes, both $f^{\rm tidal}_{lm}$, as well as the spinning $f_{lm}^{\rm S}$ expressions contain terms that scale as $f_{lm} \propto 1/\delta$. This is a removable divergence, as the Newtonian factor in the factorized waveform $h_{\ell m}^{\rm N} \propto \delta$ in Eq.~\eqref{eq:hlmFactorized}, and those terms cancel, to give a finite result. We take care of this by factorizing out the $1/\delta$ dependence in the equal-mass case in the $f_{lm}$ expression as well as the Newtonian prefactor.

We employ non-quasi-circular (NQC) corrections $N_{22}$, but only for the amplitude of the (2,2) mode and not in its phase or for any of the higher modes. NQCs are complex functions multiplied by the factorized waveform
\begin{equation}
    \label{eq:NQC}
    h_{2 2} = h_{22}^{\mathrm{F}} N_{22}(r,\Omega, p_{r_*},\bm{\lambda}),
\end{equation}
which are used in \vfivehm\ (also for the higher modes), \vfourt\ as well as \teob. They are aimed to model relevant deviations from circularity during plunge and towards the merger. It has to be noted that 
\begin{enumerate}[label=(\roman*)]
    \itemsep-3pt
    \item \vfourt\ uses the BBH NQCs one recovers from running the BBH model with the same parameters $(q, \chi_1, \chi_2)$ and applies $N_{22}^{\rm BBH}$ to the BNS waveform~\cite{Lackey:2018zvw}, while
    \item \teob\ computes BNS tuned NQCs that enforce BNS quantities at merger as fitted from NR, like frequency and its first time derivative as well as amplitude at merger~\cite{Gamba:2023mww}.
\end{enumerate}

We denote as `input values' the quantities like amplitude or frequency at merger that are extracted from NR and fit across the parameter space. In the next section we will give our reasoning for replacing the NQC corrections for the phase with a phenomenological pre-merger model tuned to NR.

\subsection{EOB pre-merger waveforms with matter effects} \label{subsec:Merger}

In the \texttt{SEOBNRv5} model, the input values and peak of the (2,2) mode are enforced at $t_{\text{peak}}^{22} = t_{\text{match}}$. We use a similar approach as is being used in the BBH model \vfivehm~\cite{Pompiliv5}, where $\tmatch$ is computed as
\begin{equation}
	\label{eq:t_attach}
	t_{\rm match} \equiv t_{\rm{ref}} - \Delta t_{22}\,,
\end{equation}
with the reference time $t_{\rm{ref}} = t_{\rm{ISCO}}$ being defined as the time at which the separation crosses $r = r_{\rm{ISCO}}$. The radius of the geodesic innermost stable circular orbit (ISCO) in Kerr spacetime $r_{\rm{ISCO}}$ is defined in Ref.~\cite{Bardeen:1972fi} and taken to be of a remnant BH with the same mass and spin one would predict from the merger of a BBH system of the same masses and spins as our compact objects, computed with NR fitting formulas from Refs.~\cite{Jimenez-Forteza:2016oae, Hofmann:2016yih}. $\Delta t_{22}=\Delta t_{22}(\bm{\lambda})$ is a calibration parameter, to be determined by comparing against NR simulations in Sec.~\ref{sec:Calibration}. 

We note that the \texttt{SEOBNRv5} family of waveform models consistently employs the remnant ISCO crossing as a reference-time in Eq.~\eqref{eq:t_attach} instead of, e.g., a peak in the orbital angular frequency as is being used in the \texttt{SEOBNRv4}\ and \teob\ families. The ISCO approach has been found to be more robust across the BBH parameter space for \texttt{SEOBNRv5HM}, as the orbital angular frequency does not necessarily peak for all systems if the resummation of the $A$-potential of the Hamiltonian is not specifically designed to enforce it. It is furthermore beneficial to employ a similar approach to determine $t_{\text{match}}$ across all \texttt{SEOBNRv5} waveform models, as this facilitates a future combination of matter effects, spin-precession and eccentricity in one \texttt{SEOBNR} waveform model.

In contrast to the BBH baseline model, we employ Eq.~\eqref{eq:t_attach} in \vfivet\ with a modified reference time 
\begin{equation}
t_{\rm{ref}}^{\rm BNS}=t_{2\rm{ISCO}}, 
\end{equation}
which is defined to be the time when the separation crosses
\begin{equation}
    r(t_{\rm{ref}}^{\rm BNS})=2\,r_{\rm{ISCO}}, 
\end{equation}
that is twice the ISCO radius of the remnant BH that one would expect if the coalescing NSs were BHs, i.e. employing the same fitting formulas as in \vfivehm. The reason for this choice is that the BBH ISCO is not necessarily crossed for BNS systems. Rather, the  
ODE integration will stop
prior to the ISCO crossing in many regions across the BNS parameter space, typically because there will be a peak in $p_{r_*}$.

This is reasonable from a physical point of view, as the NSs are likely to touch, become disrupted or strongly deformed instead of plunging into each other like BHs. The ISCO radius of a NS is, for example, expected to be comparable to its physical radius $r_{\rm{ISCO}} \approx r_{\rm{NS}}$, but can be larger or smaller dependent on the NS's EoS and resulting compactness. Consequently, for different BNS systems, the ISCO crossing of the EOB system and the NSs touching each other and merging could occur in any order. Changing the arbitrary reference point to the earlier time of the $r_{\rm ref} = 2\,r_{\rm{ISCO}}$ crossing that is reached across the relevant parameter space therefore gives us more control on $t_{\rm match}$. This control is of importance for a more accurate modeling of the time of merger and the pre-merger waveform, as we discuss below, but also to increase the robustness of our EOB model under small perturbations of the intrinsic parameters.

\begin{figure}[t]
    \centering
   \includegraphics[width=0.97\linewidth]{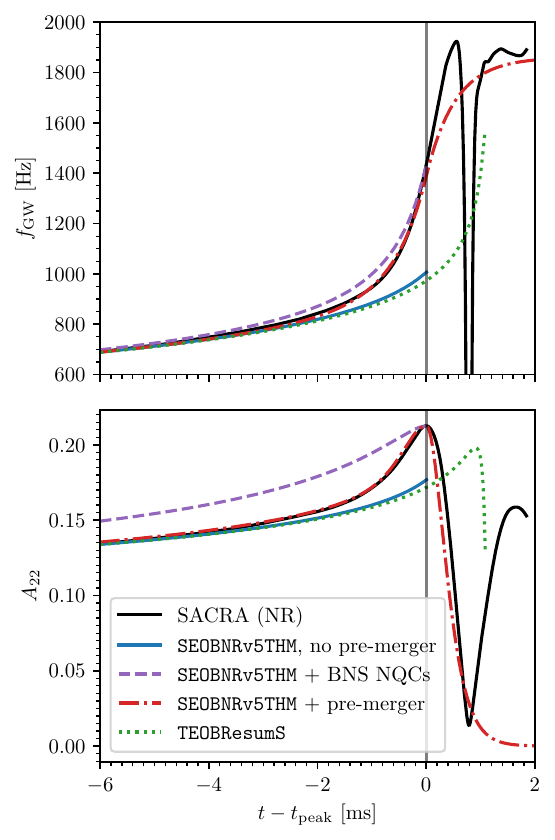}
   \caption{The frequency-time (top) and amplitude-time evolution (bottom) of different pre-merger prescriptions compared to an exemplary SACRA simulation with EoS 15H for a $(1.25 + 1.25)\ M_{\odot}$ BNS. The NR waveform is shown as a solid, black line and the \vfivet\ factorized waveform from Eq.~\eqref{eq:hlmFactorized} without application of any pre-merger model as a blue, solid line. The application of BNS-tuned NQCs on top of the factorized waveform is shown as a purple, dashed line and the phenomenological pre-merger and tapered model we use in \vfivet\ as outlined in this section as a dot-dashed, red line. We further show \teob\ as a dotted, green line and indicate the time of merger of the NR simulation $t_{\rm peak}=\tmatch$ with a vertical line.}
   \label{fig:frequency}
\end{figure}

Regarding the waveform prior to the peak emission, we opt to take a different approach to what has been presented in the literature so far to model the non-linear matter and gravity effects close to merger. Instead of BNS-tuned NQCs as have been developed and used in \teob~\cite{Gamba:2023mww}, we develop a phenomenological model for the phase and amplitude close to merger.
Our reason for this is that on some parts of the parameter space, BNS-tuned NQCs can lead to a dephasing and overestimated amplitude of the waveform towards merger in comparison to NR.
This can be seen in the top panel of Fig.~\ref{fig:frequency}, where we show the frequency-time evolution around merger for the exemplary SACRA simulation of a $(1.25 + 1.25)\ M_{\odot}$ BNS for different pre-merger models. We remind that an incorrect frequency evolution translates into a dephasing over time in comparison to NR.

When one
uses the factorized waveform from Eq.~\eqref{eq:hlmFactorized} (without application of NQCs or a pre-merger modeling and only up to $\tmatch$), the frequency evolution of the EOB model is too `soft' in comparison with the NR one towards merger, i.e. the frequency does not increase fast enough towards merger. We find this result across all NR simulations that we have investigated and one could argue that this might be due to friction when the NSs touch and exchange matter~\cite{Kiuchi:2017pte}, as well as higher order PN effects. Notably, there also exists the possibility that the development of tidal spin, even for non-spinning NSs, might `harden' the frequency evolution~\cite{Yu:2025}. All of these phenomena are however yet to be modeled analytically in an efficient way. For example, a full evolution of the NS mode amplitudes during $f$-mode resonance (in contrast to our effective description) would increase the waveform generation speed substantially~\cite{Steinhoff:2016rfi,Yu:2025}.

Despite the fact that \teob\ has BNS-tuned NQCs, its time of merger is not calibrated to NR, and it therefore enforces the NR input values $\sim 1$ ms too late for this particular simulation. As a consequence, one can see that its frequency evolution is also too `soft', highlighting the importance of an accurate prediction of the merger time. This is also observable later in Sec.~\ref{subsec:Phasing}, where \teob\ can accumulate dephasing towards merger for some NR simulations.

In contrast, note that the application of BNS-tuned NQCs to the \vfivet\ waveform inspired by the \teob\ approach makes the frequency-evolution too `hard', i.e. the frequency is too high during the pre-merger, in particular, as the NQCs are enforced at the correct time of merger and enforce the correct frequency and first frequency derivative at merger. For this work, we have tried out different functional forms for the NQCs, by employing the same NQCs as are being used in \vfivehm~\cite{Pompiliv5}
\begin{equation}
    \begin{split}
        \label{eq:nqc}
        N_{\ell m} = &\left[1+\frac{\hat{p}_{r_*}^2}{(r \Omega)^2}\left(a_1^{h_{\ell m}}+\frac{a_2^{h_{\ell m}}}{\hat{r}}\right)\right] \\ &\times \exp \left[i\left(b_1^{h_{\ell m}} \frac{\hat{p}_{r_*}}{r \Omega}+b_2^{h_{\ell m}} \frac{\hat{p}_{r_*}^3}{r \Omega}\right)\right];
    \end{split}
 \end{equation}
as well as the same form with modified powers of $\hat{r}=r/M$ and $\hat{p}_{r_*}=p_{r_*}/\mu$ to `harden' the amplitude and frequency evolution; and by testing the different NQC-prescription employed by \teob, given in Eqs. (33)--(37) of Ref.~\cite{Gamba:2023mww} instead. We have found no satisfactory results across the parameter space covered by NR when employing NQCs with the input parameters enforced at the NR time of merger. Rather, we find that some of the NR waveforms will always be modeled more inaccurately due to the application of NQCs as described above.

We argue that this can be expected as the functional form of the NQCs is chosen to model deviations from quasi-circularity during the plunge. As a BNS waveform is largely driven by matter-effects around merger instead, we suggest that they should not blindly be used to enforce BNS input values. Very similar observations also hold for the amplitude towards merger, as one can see in the bottom panel of Fig.~\ref{fig:frequency}. Note that for both panels we employ NQCs that enforce the correct frequency, first frequency derivative, amplitude, and a peak of the amplitude at $\tmatch$ through Eq.~\eqref{eq:nqc}.

Although it might be worthwhile to study how matter effects should modify the NQC functional form, starting from theoretical considerations in the future, in this work, we instead suggest a different, more agnostic approach. It is motivated by the idea of applying a simple exponential function to the frequency and amplitude of \vfivet, which accurately models the frequency-evolution as well as the amplitude-evolution of a given BNS system up to merger  across the entire parameter space covered by NR, as we show in detail in Sec.~\ref{sec:Validation}.

The (2,2) mode containing the pre-merger modeling and post-merger tapering has the form
\begin{equation}
    \label{eq:h22_modeled}
    h_{22}^\text{pre-merge}(t)=A_{22}^\mathrm{phen}(t) W(t) e^{-i\phi^\mathrm{phen}_{22}(t)},
\end{equation}
where the phenomenologically modeled phase $\phi^\mathrm{phen}_{22}(t)$ and amplitude $A_{22}^\mathrm{phen}(t)$ are defined below in Eqs.~\eqref{eq:phaseboost} and \eqref{eq:Aboost} respectively, and $W(t)$ is a Planck window taper, inspired by the tapering of \vfourt~\cite{Lackey:2018zvw}
\begin{equation}
    W(t)=\left[1 + \exp\left(\frac{t - t_{\rm match}}{\tau_W}\right)\right]^{-1}
    \label{eq:transition}
\end{equation}
which in our case is centred at $\tmatch$ and with width $\tau_W$ that is defined in Eq.~\eqref{eq:amplitude} below.

\begin{table}[t] 
    \renewcommand*{\arraystretch}{1.4}
    \caption{\label{tab:free_params} A summary of the free parameters and functions used in the phenomenological pre-merger model presented in this section. The free parameters are heuristically set to recover agreement with NR, while some parameters are constrained to recover input values, continuity in the first derivative throughout the merger, or consistency of the assumption of positive decay times $\tau > 0$.}
    \begin{ruledtabular}
        \begin{tabular}{cccc}
         Function & Definition & Parameter & Constraint
         \\ \hline 
         \multirow{2}{*}{\centering Phase boost $\phi_{22}^\mathrm{phen}(t)$} & \multirow{2}{*}{\centering Eq.~\eqref{eq:phaseboost}} & $B_\omega$ & input value \\ 
          &  & $\tau_\omega$ & free \\ 
          \multirow{2}{*}{\centering Post-merger phase } & \multirow{2}{*}{\centering Eq.~\eqref{eq:phasetaper}} & $\omega^{\rm asymp}_{22}$ & free \\ 
         & & $(\Delta \omega,\ \tau_{\rm taper})$ & continuity \\ \hline
         Tapering $W(t)$ & Eq.~\eqref{eq:transition} & $\tau_W$ & merger time \\ 
         NQC $N_{22}$ & Eq.~\eqref{eq:NQC_v5} & $a_1^{h_{22}}$ & input value \\
         \multirow{2}{*}{\centering Amp. boost $A^\mathrm{phen}(t)$} & \multirow{2}{*}{\centering Eq.~\eqref{eq:Aboost}} & $B_A=1$ & input value \\ 
          &  & $\tau_A$ & consistency \\ 
         Residual freedom & Eq.~\eqref{eq:amplitude} & $\tau_{\rm free}$ & free \\
            \end{tabular}
    \end{ruledtabular}
\end{table}

Starting from the factorized waveform as given by Eq.~\eqref{eq:hlmFactorized}
\begin{equation}
    h_{22}^{\rm F}(t)=A_{22}(t) e^{-i\phi_{22}(t)}
\end{equation}
we want to boost the angular frequency $\omega_{22}=\dot{\phi}_{22}$  as
\begin{equation} \label{eq:boost}
    \omega_{22}^\mathrm{phen}(t) = \omega_{22}(t) \beta(t|B_\omega,\tau_\omega,t_\mathrm{match}),
\end{equation}
by an exponential ansatz of the form
\begin{equation} \label{eq:boost_function}
    \beta(t|B,\tau,t_\mathrm{match}) = 1+B \exp \left(\frac{t-t_\mathrm{match}}{\tau}\right).
\end{equation}
This ansatz keeps the frequency-evolution based on the EOB waveform, but exponentially `hardens' it towards merger as depicted in Fig.~\ref{fig:frequency} and suggested by NR. Below we apply the same ansatz to the amplitude as well, but with different coefficients $B$ and $\tau$.

To obtain the phase of the pre-merger waveform $\phi_{22}^\mathrm{phen}(t)$, we integrate the angular frequency for pre-merger times $t\leq \tmatch$ 
\begin{equation}
    \begin{split}
        \phi_{22}^\mathrm{phen}(t) = &\ \phi_{22}(t_{\rm trans}) \\ &+ \int_{t_{\rm trans}}^t\omega_{22}(t') \left(1+B_\omega e^{(t'-\tmatch)/\tau_\omega}\right) \d t',
    \end{split}
\end{equation}
where $t_{\rm trans}$ is defined in Eq.~\eqref{eq:ttrans} below and marks the start of the pre-merger waveform. This integral can be written in the form
\begin{equation}
    \begin{split} \label{eq:phaseboost}
        \phi_{22}^\mathrm{phen}(t) = \, &\phi_{22}(t) + B_\omega \phi_{22}(t') e^{(t'-\tmatch)/\tau_\omega}\bigg|_{t_{\rm trans}}^t \\ 
        &- \frac{B_\omega}{\tau_\omega}\int_{t_{\rm trans}}^t\phi_{22}(t') e^{(t'-\tmatch)/\tau_\omega} \,\d t',
    \end{split}
\end{equation}
where the integral in the last line is performed numerically through a cumulative sum of the integrand on a uniform sample grid of $10^3$ points between $t_{\rm trans}$ and $\tmatch$, i.e., through the rectangle rule. Given the phenomenological nature of our pre-merger approach, we have decided against the usage of a more accurate, but computationally more intensive higher-order integration method.

We use fits for the BNS-tuned input values of the merger frequency $\omega_{22}^\mathrm{NR}$ and merger amplitude $A_{22}^\mathrm{NR}$ from Ref.~\cite{Breschi:2022}. From Eqs.~\eqref{eq:boost} and \eqref{eq:boost_function}, one can see that
\begin{equation} \label{eq:Bomega}
    B_\omega \equiv \frac{\omega_{22}^\mathrm{NR}}{\omega_{22}} - 1,
\end{equation}
enforces the correct input value frequency at $\tmatch$. We further choose $\tau_\omega=1.75\,\pi/\omega_{22}^\mathrm{NR}$ as we find it produces a good match with NR across all simulations and can be interpreted as the boost taking effect over the last 1.75 orbital cycles.

We simply continue the waveform frequency from the merger onwards to an asymptotic value of $\omega_{22}^{\rm asymp}=1.2\, \omega_{22}^\mathrm{NR}$ by setting
\begin{equation} \label{eq:phasetaper}
    \omega_{22}^{\rm phen}(t) = \omega_{22}^{\rm asymp} - \Delta \omega e^{-(t-\tmatch)/\tau_{\rm taper}}
\end{equation}
for $t>\tmatch$ and by tuning $\Delta \omega$ and $\tau_{\rm taper}$ such that the first frequency derivative is kept continuous. One finds the tapered phase by analytically integrating this equation.
We highlight that a continuous first frequency derivative was not enforced in the tapering of \vfourt~\cite{Lackey:2018zvw}, which has a discontinuous $\dot{\omega}_{22}$ at merger.

Regarding the pre-merger model for the amplitude $A_{22}^\mathrm{phen}(t)$, we want to 
\begin{enumerate}[label=(\roman*)]
    \itemsep-3pt
    \item enforce the correct amplitude at merger,
    \item the peak to occur at $\tmatch$, and
    \item to taper the waveform to zero in a continuously differentiable way
\end{enumerate}
while recovering the amplitude-evolution as given by NR faithfully across all simulations used in this work. Our approach also has to produce a robust model without waveform generation errors, unintended artefacts or pathologies across the parameter space.

To this end, we have decided to employ one NQC that enforces the correct amplitude at merger and then using the boost function from Eq.~\eqref{eq:boost} to enforce the NR calibrated merger amplitude and time of merger in the presence of the tapering $W(t)$, equalling one function per constraint.

We employ an amplitude NQC inherited from the NQCs as used in \vfivehm, see Eq.~\eqref{eq:nqc}
\begin{equation} \label{eq:NQC_v5}
    N_{22}(t) = 1+\frac{p_{r_*}^2}{(r \Omega)^2}a_1^{h_{22}}
\end{equation}
with one free parameter $a_1^{h_{22}}$ tuned such that 
\begin{equation} \label{eq:amp_IV}
    A_{22}(\tmatch) N_{22}(\tmatch) = A_{22}^\mathrm{NR}
\end{equation}
with $A_{22}^\mathrm{NR}$ from Ref.~\cite{Breschi:2022}.

We then boost the amplitude by applying the ansatz from Eq.~\eqref{eq:boost} up to $\tmatch$. After the merger, we linearly continue the resulting amplitude based on the amplitude after application of the NQC and boost. Prior to windowing, we therefore arrive at a continuously differentiable, phenomenological amplitude
\begin{align}
    \label{eq:Aboost}
    A_{22}^\mathrm{phen}(t) = \begin{cases}
        \begin{aligned}[t]
            & A_{22}(t) \, N_{22}(t) \, \beta(t|B_A,\tau_A,t_\mathrm{match})\\
            & \qquad\qquad\qquad\qquad \mathrm{for}\ t \leq t_\mathrm{match},
        \end{aligned} \\\\
        \begin{aligned}[t]
            & A_{22}^\mathrm{phen}(t_\mathrm{match}) \\ &+ \dot{A}_{22}^\mathrm{phen}(t_\mathrm{match}) (t-\tmatch)\\
            & \qquad\qquad\qquad\qquad\mathrm{for}\ t > t_\mathrm{match},
        \end{aligned}
    \end{cases}
\end{align}
with two free parameters $B_A$ and $\tau_A$.

By inserting Eqs.~\eqref{eq:h22_modeled}--\eqref{eq:Aboost} into each other one can check that the correct amplitude at merger $\tmatch$ requires $B_A=1$. By differencing those equations at $t=\tmatch$ one can also see that defining
\begin{equation} \label{eq:amplitude}
    \begin{split}
        \tau_A &\equiv \frac{\tau_{\rm free}}{1+\Big[1+\mathrm{sgn} (\mathcal{A})\Big]\tau\mathcal{A}  }, \\
        \tau_W &\equiv \frac{\tau_{\rm free}}{1+\Big[-1+\mathrm{sgn} (\mathcal{A})\Big]\tau\mathcal{A}  },
    \end{split}
\end{equation}
yields positive decay times $(\tau_A,\,\tau_W)>0$ independently on the behavior of the original amplitude $A_{22}$ as well as a correct time of merger.
The free parameter $\tau_{\rm free}=1.5\,\tau_\omega$ is set heuristically to faithfully model the pre-merger amplitude across all NR simulations we have studied. We have further used the quantity
\begin{equation}
    \mathcal{A} = \left.\frac{\d \ln \left(A_{22}^{\rm phen }\right)}{\d t}\right|_\tmatch=\frac{\dot{A}_{22}^\mathrm{phen}(t_\mathrm{match})}{A_{22}^\mathrm{phen}(t_\mathrm{match})}.
\end{equation}

Through this procedure we get a phenomenological pre-merger model, that proves to be accurate and robust across the whole parameter space as we show in Sec.~\ref{sec:Validation}. We summarize all parameters which we have introduced for this purpose in Table \ref{tab:free_params}.

As a last step, we set the beginning of our pre-merger-tapered waveform $t_{\rm trans}$ from Eq.~\eqref{eq:h_match} as
\begin{equation} \label{eq:ttrans}
    t_{\rm trans} = \tmatch - 15 \max \left(\tau_\omega,\,\tau_A,\, \tau_W \right)
\end{equation}
the maximum of all the exponential decay times.

For low tidal deformabilities $\kappa_2^T < 20$ we smoothly transition the matching time $t_{\rm match}^{\rm BNS}$ of Eq.~\eqref{eq:t_attach} and the input values $\omega_{22}^\mathrm{NR}$ and $A_{22}^{\rm NR}$ to the BBH quantities as used in \vfivehm\ through the smooth transition function
\begin{equation} \label{eq:SmoothTransition}
    \sigma(x) = \begin{cases}
        0  \quad & \mathrm{for}\ x \le 0\,, \\
        \frac{\exp \left(\frac{-1}{x}\right)}{\exp \left(\frac{-1}{x}\right) + \exp \left(\frac{-1}{1-x}\right)} \quad & \mathrm{for}\ 0 < x < 1\,,\\
        1 \quad & \mathrm{for}\ x \geq 1\,,
    \end{cases}
\end{equation}
rescaled to be applied on the interval $\kappa_2^T \in [10,20]$. This way, we recover the BBH baseline to the $10^{-4}$ to $10^{-3}$ mismatch level in the absence of matter as we discuss further in Sec.~\ref{subsec:Mismatches}. There is however no exact match to BBH waveforms, due to the absence of a merger-ringdown model as well as the different pre-merger modeling.

As it was not the scope of this work to model the merger waveform for the higher modes, we simply taper them similarly to the \vfourt\ approach in an unmodeled way. We describe the tapering for the subdominant modes in Appendix~\ref{app:tapering} and leave the development of a similar but faster approach for the higher modes as future work.

The reasons for this choice are that higher order modes are suppressed for binaries of similar masses $q \lesssim 3$ like BNS systems~\cite{Cotesta:2018fcv}, and that the mode frequencies scale approximately as $f_{lm} \simeq (m/2)\ f_{22}$~\cite{Ujevic:2022qle}. High $m$ modes are therefore currently outside of the detector bands close to merger, and a correct pre-merger model of the (2,2) mode is of greater importance. Additionally, only very few BNS simulations have a decomposition of the NR waveform into spherical harmonics for us to compare or calibrate against. 
We therefore leave the extension of our model for the merger of all higher-modes as future work.

We however perform a validation study of the higher modes compared to NR in Appendix~\ref{app:hm}, which shows that the current model is nonetheless able to generate accurate higher modes until close to merger.

\subsection{Waveform generation speed} \label{subsec:PA}

We use the same procedure as is being used in \vfivehm\ for the PA approximation, see Eqs. (17)--(20) of Ref.~\cite{Pompiliv5}, which allows us to speed up the computation of very long waveforms, as derived in Refs.~\cite{Damour:2012ky,Nagar:2018gnk}. This technique has previously been used extensively in the \teob\ family of models~\cite{Rettegno:2019tzh,Nagar:2020pcj,Riemenschneider:2021ppj,Gamba:2021ydi}.

\begin{figure}[t]
    \centering
   \includegraphics[width=.99\linewidth]{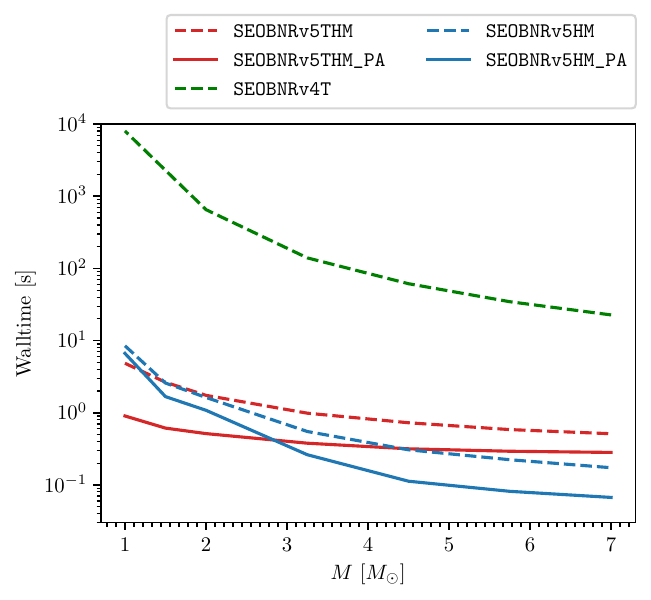}
   \caption{Benchmarks for \vfivet, our baseline \vfivehm\ and the predecessor approximant \vfourt\ for a BNS system $(q=1.5,\,\Lambda_1=700,\,\Lambda_2=800,\,\chi_1=0.1,\,\chi_2=0.2)$ starting from $20$ Hz where we only compute the $(2,2)$ mode. For the \texttt{SEOBNRv5} models, we also show the benchmarks if the PA approximation is applied as solid lines. We call both tidal approximants at a sampling rate of $f_{\rm sample}=8192$ Hz and note that for low total-mass systems, \vfivehm, which is run with $\Lambda_1=\Lambda_2=0$, needs a higher sampling rate to resolve the ringdown, which is why \vfivet\ can become faster than it's baseline. These benchmarks were run on an Apple M2 processor.}
   \label{fig:timing}
\end{figure}

The idea is to obtain and solve explicit algebraic equations for the momenta over a grid of radii covering the inspiral, by perturbatively solving for the deviations from circularity $p_{r_*}=0$. This way one can circumvent the explicit ODE integration to get an accurate and fast waveform. Like \vfivehm, we employ the PA approximation at the $8^{\rm th}$ order.

It has to be noted that the usage of the PA approximation, our changes in the $\keff$ explained around Eq.~\eqref{eq:1}, and our code optimizations have increased the waveform generation speed of \vfivet\ by two to four orders of magnitude in comparison to \vfourt\ for systems with $M \geq 1\, M_\odot$, which can be seen in Fig.~\ref{fig:timing}. This allows us to use the current model for the PE of BNS events, as we show in Sec.~\ref{sec:PE}. 

We highlight however that the speed of the \vfourt\ model has been greatly increased through the development of the frequency-domain \texttt{SEOBNRv4T\_surrogate} in Ref.~\cite{Lackey:2018zvw}. This has made the \vfourt\ model even faster than \vfivet, and further allows the application of \vfourt\ in PE. As the surrogate has been built in frequency-domain, its waveform generation speed is however not directly comparable to time-domain models, in particular for very long waveforms. This is also why we omit the timing comparison to the surrogate and \texttt{NRTidalv3} models from Fig.~\ref{fig:timing}.

For reference, the generation of a single frequency-domain \texttt{SEOBNRv4T\_surrogate} waveform from 20 to 4096 Hz with frequency sampling rate $\Delta f = 1/1024$~Hz takes on the order of 0.1\,s, largely independent of total mass. The same holds approximately true for an \texttt{IMRPhenomXAS\_NRTidalv3} waveform. In contrast, the generation of an EOB waveform in frequency-domain is slowed down by an additional discrete Fourier transform. Compared to a surrogate, an EOB model has however the advantage of extensibility, i.e. new analytical or numerical information can be easily included as it becomes available. Furthermore, PE with \vfivet\ is only twice as expensive as PE with \texttt{IMRPhenomXAS\_NRTidalv3} when using \texttt{parallel\ bilby}, even though the waveform generation cost is a few times higher, as we discuss in Sec.~\ref{sec:PE}. Note that this does not necessarily apply to all PE codes, in particular PE codes employing multi-banding~\cite{Vinciguerra_2017,Ashton_2019,Dax:2024}.

\section{Calibration to numerical relativity waveforms} \label{sec:Calibration}

An advantage of the time-domain EOB models over frequency-domain models is the access to the binary's separation $r(t)$, enabling improved estimates of the merger time via Eq.~\eqref{eq:t_attach}.

Accurately predicting the merger time is crucial as an observable, potentially detectable by future facilities like ET~\cite{Breschi:2022}. Of more importance is however its use in refining the pre-merger waveform modeling as described in Sec.~\ref{subsec:Waveform}, as the matching time $\tmatch$ significantly affects pre-merger frequency and amplitude evolution.

For the calibration and validation of \vfivet, 45 NR waveforms simulated by the SACRA code~\cite{Kiuchi:2017pte, Kawaguchi:2018gvj, Kiuchi:2019kzt}, and 19 waveforms simulated with the BAM code~\cite{Dietrich:2018phi, Ujevic:2022qle, Gonzalez:2022mgo} were employed. We provide more detailed information about the individual simulations in Appendix~\ref{app:NR_calib}.
We use NR input values from Ref.~\cite{Breschi:2022} for the peak amplitude and frequency. The only quantity of \vfivet\ calibrated in this work is thus the time of merger $\dt$ given in Eq.~\eqref{eq:t_attach}. We do however also employ NR fits for the BBH ISCO of the BBH remnant used in the computation of $t_{\rm ref}$, as explained below Eq.~\eqref{eq:t_attach}, as well as for the BNS input values as used in our phenomenological pre-merger model in Sec.~\ref{subsec:Merger}.

For all simulations, we align our model at the beginning of the respective NR simulation by minimizing the $(2,2)$ mode phase difference between a given NR simulation and our model
\begin{equation} \label{eq:alignment}
    \min_{(t_0,\,\phi_0)}\ \int_{t_{\rm start}}^{t_{\rm end}} |\phi_{22}^{\rm NR}(t) - \phi_{22}^{\rm EOB}(t - t_0) - \phi_0|\, \d t,
\end{equation}
where $\phi_{22} (t)=\arg [h_{22}(t)]$. We align SACRA waveforms, which tend to be longer than the BAM counterparts, over 7 waveform cycles and BAM waveforms over 5 cycles, setting $t_{\rm start}$ at a point where the junk radiation has decayed. We then extract the $\dt$ of Eq.~\eqref{eq:t_attach} that matches the merger of the NR simulation to $\tmatch$ of \vfivet.

\begin{figure}[t]
    \centering
   \includegraphics[width=\linewidth]{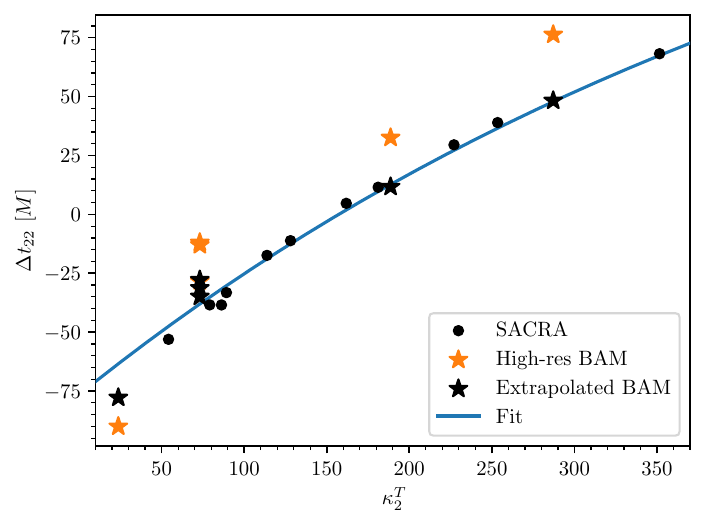}\hfill
   \caption{The resulting $\dt$ for the equal-mass NR simulations prior to and after the extrapolation to a common resolution. We show the recovered $\dt$ that matches the merger of the NR simulation to $t_\mathrm{attach}$ from Eq.~\eqref{eq:t_attach} for all used equal-mass simulations, given their respective quadrupolar tidal coupling constant $\kappa_2^T$. We indicate with black dots the employed SACRA simulations and with stars the BAM simulations. The originally recovered $\dt$ values for the highest-resolution BAM simulation are shown in orange, and the $\dt$ extrapolated to SACRA resolution in black. We also indicate the resulting fit from Eq.~\eqref{eq:t_fit} as a blue line.}
   \label{fig:eq_mass}
\end{figure}

We find that the time of merger strongly depends on the employed resolution of the NR simulation, where for the convergent BAM simulations, which have lower total resolution than the SACRA simulations, a higher resolution typically corresponds to a later time of merger.

To have data for $\dt$ that behaves continuously across the parameter space and can be fitted, we therefore have to extrapolate the resolution dependent $\dt$ to a common resolution across the NR simulations used in this work. We illustrate this in Fig.~\ref{fig:eq_mass}. We observe that the extrapolation of the merger times for BAM, as described below, makes $\dt$ continuous and fittable across the parameter space. Note that we fit $\dt$ across tidal parameters and mass-ratio simultaneously, which yields the slight mismatch for the low tides, equal-mass case in the Fig.~\ref{fig:eq_mass}.

\begin{figure}[t]
    \centering
   \includegraphics[width=\linewidth]{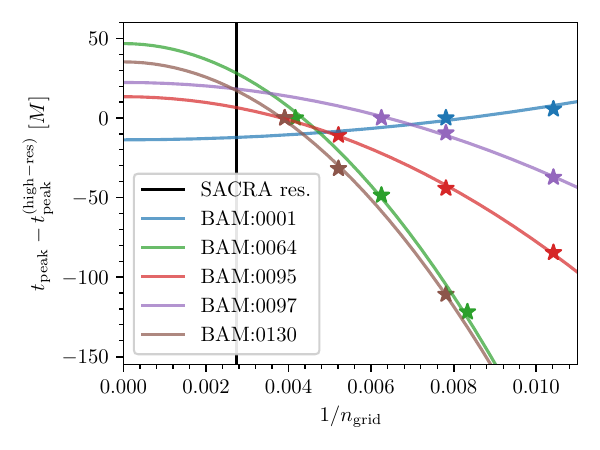}\hfill
   \caption{Extrapolation of the merger time for some BAM simulations of different resolutions $1/n_{\rm grid}$. We show the difference between the time of merger $t_{\rm peak}$ for runs at different resolutions and the time of merger of the respective highest resolution run $t_{\rm peak}^{\text{(high-res)}}$ of a given system as stars of the same color. We also include the respective quadratic fit of the data as a solid line. The resolution employed by SACRA simulations is indicated by a black vertical line, from where we extract $\Delta t_{\rm peak}^{\rm (extrap)}$ as the value of the quadratic fit at the SACRA resolution, see Eq.~\eqref{eq:tpeakNR}.}
   \label{fig:BAM_t_peak}
\end{figure}

As all SACRA simulations employ a uniform highest grid spacing of $n_{\rm grid}=182$, while the BAM simulations have lower resolutions\footnote{The way in which the SACRA and BAM simulations label their grid is different in such a way, that the SACRA grid number is twice the one of BAM.}, but are convergent at a clear convergence order, we have opted to extrapolate the BAM $\dt$ to a value that corresponds to the SACRA resolution as depicted in Fig.~\ref{fig:BAM_t_peak}.

For this matter we extract the time of merger among the four highest resolution BAM simulations\footnote{If only three runs are available or the fourth run has a too low resolution, we opt to only use the three highest resolution runs for the fit.}, and fit the resolution-dependent merger time as
\begin{equation}
    t_{\rm peak}(1/n_{\rm grid})=a/n_{\rm grid}^2 + b.
\end{equation} 
We extract the value of $t_{\rm peak}(1/n_{\rm SACRA})$ and then calibrate as an extension of Eq.~\eqref{eq:t_attach}
\begin{equation}
	t_{\rm peak}^{22} + \Delta t_{\rm peak}^{\rm (extrap)} = t_{\rm{ref}} - \Delta t_{22}\,,
\end{equation}
where 
\begin{equation}
    \label{eq:tpeakNR}
    \Delta t_{\rm peak}^{\rm (extrap)} = t_{\rm peak}(1/n_{\rm SACRA}) - t_{\rm peak}^{\rm (high-res)}.
\end{equation}

\begin{figure}[t]
    \centering
   \includegraphics[width=\linewidth]{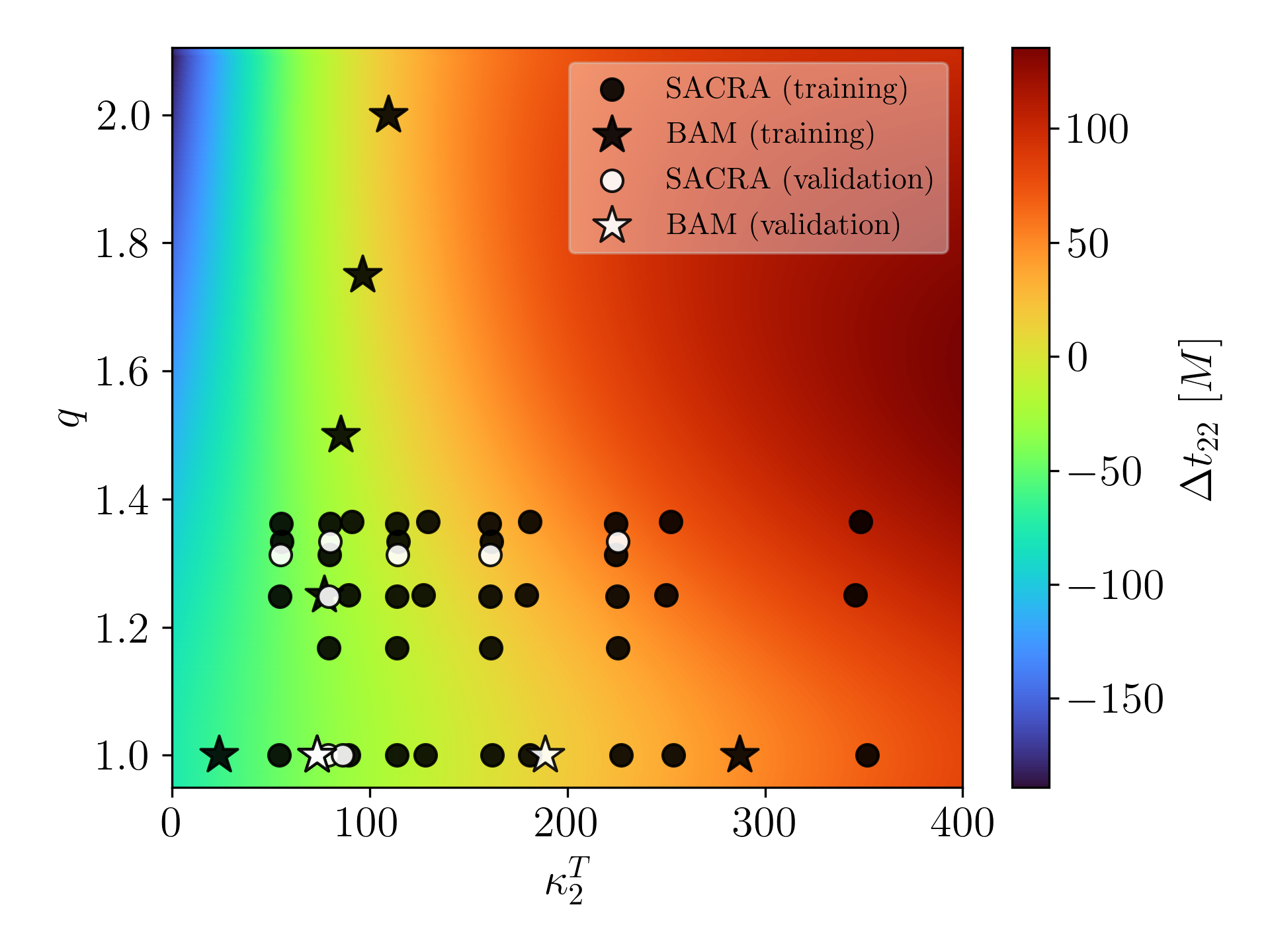}\hfill
   \caption{$\dt$ fit across the parameter space of the employed simulations. In the background of the image we show the result of our fit, while the data points show the respective mass-ratio $q$ and quadrupolar tidal coupling constant $\kappa_2^T$ of the employed non-spinning simulations. We indicate SACRA simulations as circles and BAM simulations as stars. Points with a black (white) interior have been used for fitting (validation).}
   \label{fig:fit_parameterspace}
\end{figure}

We have also explored the possibility of extrapolating both SACRA and BAM to infinite resolution, but as the SACRA waveforms are not clearly convergent, we have found their merger times to not be clearly convergent. By trying to extrapolate the SACRA merger times, one therefore introduces noise and the resulting $\dt$ becomes less continuous and therefore fittable. Additionally, the agreement with NR in Sec.~\ref{sec:Validation} worsens through this procedure, as we are not calibrating to the finite-resolution NR waveforms anymore but their extrapolations instead.

This is also the reason why in Sec.~\ref{subsec:Mismatches} our mismatches compared to the SACRA simulations are better than our mismatches to BAM simulations, as we are not calibrating to the highest resolution BAM simulations directly.

The resulting values of $\dt$ at SACRA resolution are then indeed continuous across the parameter space, as one can see in Fig.~\ref{fig:fit_parameterspace}, where our fit is shown in the background together with the parameters used for training and validation. We employ the fitting function
\begin{equation}\label{eq:t_fit}
        \dt(X,\kappa_2^T) = a - b\,e^{ -c\,\kappa_2^T (1 + dX +eX^2) + f} - gX ,
\end{equation}
where $X = 1-4\,\nu \in [0,1]$ and use 80\% of the simulations for the fit while leaving 20\% of the simulations for validation. The resulting fitting parameters are shown in Table~\ref{tab:fit}. We choose the above functional form to ensure an asymptote $a-gX$ for tidal deformabilities outside of the calibrated region $\kappa_2^T \gg 400$, which is suggested by the data and furthermore keeps our $\tmatch$ bound across the whole parameter-space as $\dt \in [-970,\, 186]\, M$ for any system, which prevents waveform generation failures. The quality of the fit is expressed in Table~\ref{tab:fit_quality}.

\begin{table}[t] 
    \renewcommand*{\arraystretch}{1.4}
    \caption{\label{tab:fit} The fitting parameters for $\dt$ in Eq.~\eqref{eq:t_fit}.}
    \begin{ruledtabular}
        \begin{tabular}{cccc}
        $a$ & $b$ & $c$ & $d$ \\
        215.5214 & 177.0887 & $1.932128 \times 10^{-3}$ & 31.37043 \\ \hline
        $e$ & $f$ & $g$ & - \\
        59.45997 & 0.500273 & 892.9965 & - \\
            \end{tabular}
    \end{ruledtabular}
\end{table}

\begin{table}[t] 
    \renewcommand*{\arraystretch}{1.4}
    \caption{\label{tab:fit_quality} The quality of the \vfivet\ $\dt$ calibration and parameters for the fit. The table lists the mean absolute error and $90\%$ percentile of the absolute error of the (extrapolated) $\tmatch$ in units of total mass and for a $(1.4+1.4)$ $\MSun$ BNS system in ms. We also compare the absolute error of the \vfourt\ merger time after alignment as a comparison of a model without calibration.}
    \begin{ruledtabular}
        \begin{tabular}{ccccc}
         & \multicolumn{2}{c}{\centering Mean absolute error} & \multicolumn{2}{c}{\centering 90 \% absolute error} \\
         & [$M$] & [ms] & [$M$] & [ms] \\
        \hline
        Fitting & 4.87 & 0.0672 & 10.03 & 0.138 \\ 
        Validation & 3.74 & 0.0515 & 6.74 & 0.0929 \\ 
        \vfourt & 10.33 & 0.1425 & 13.83 & 0.191 \\
            \end{tabular}
    \end{ruledtabular}
\end{table}

We did not calibrate $\dt$ to spinning BNS systems, as there are not enough NR simulations to cover a large portion of the parameter space, but have instead opted to use $\dt + \Delta t_{22}^{\rm spin}$ from the baseline model \vfivehm, see Eqs. (73) and (80) of Ref.~\cite{Pompiliv5} to include the effects of non-precessing, spinning systems on the merger time. As $\tmatch$ is computed in a linear way in both cases, this should include all point-mass spin-effects that augment the merger time for BNS systems.

\section{Validation of tidal waveform approximants} \label{sec:Validation}

To assess the impact of the changes and improvements introduced in the \vfivet\ waveform model, we compare it to 
the set of NR simulations described in Sec.~\ref{sec:Calibration}, and to other state-of-the-art BNS approximants, namely the predecessor model \vfourt, as well as \seobnrvfiveromnrtidalvthree, and \teobg. We also assess the accuracy of \vfivet\ when modeling BBH systems by comparing against \vfivehm.

We do so by showing a select few time-domain phasings of BNS simulations with the available waveform approximants in Sec.~\ref{subsec:Phasing}, by performing unfaithfulness computations to NR simulations in Sec.~\ref{subsec:NR}, as well as between different waveform approximants across the parameter space of interest in Sec.~\ref{subsec:Mismatches}.

\subsection{Phasing comparison} \label{subsec:Phasing}

\begin{figure*}[t]
    \centering
   \includegraphics[width=0.5\linewidth]{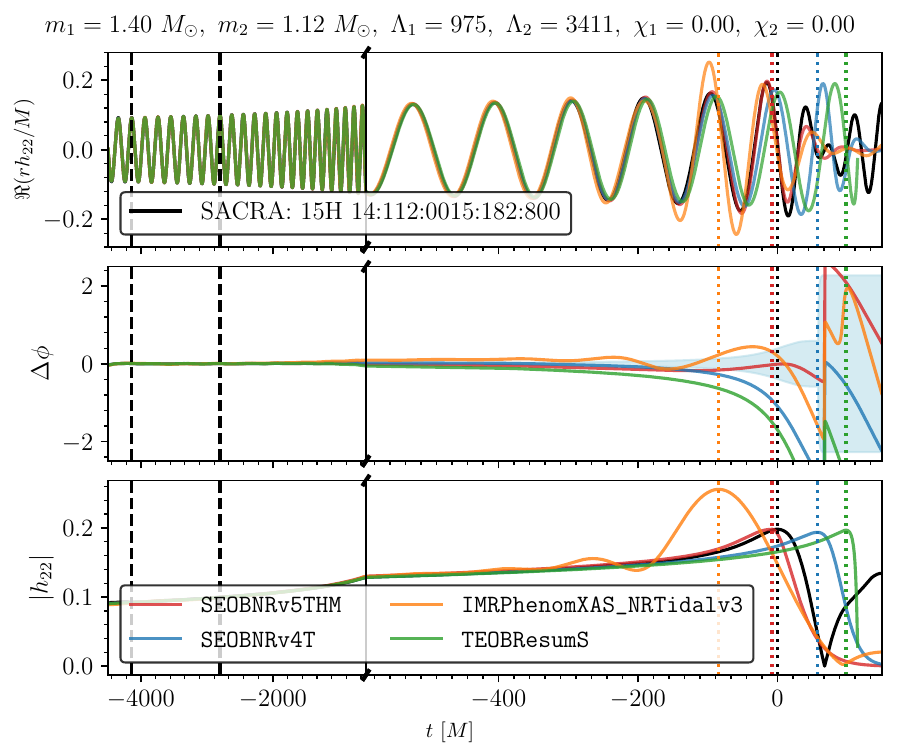}\hfill
   \includegraphics[width=0.5\linewidth]{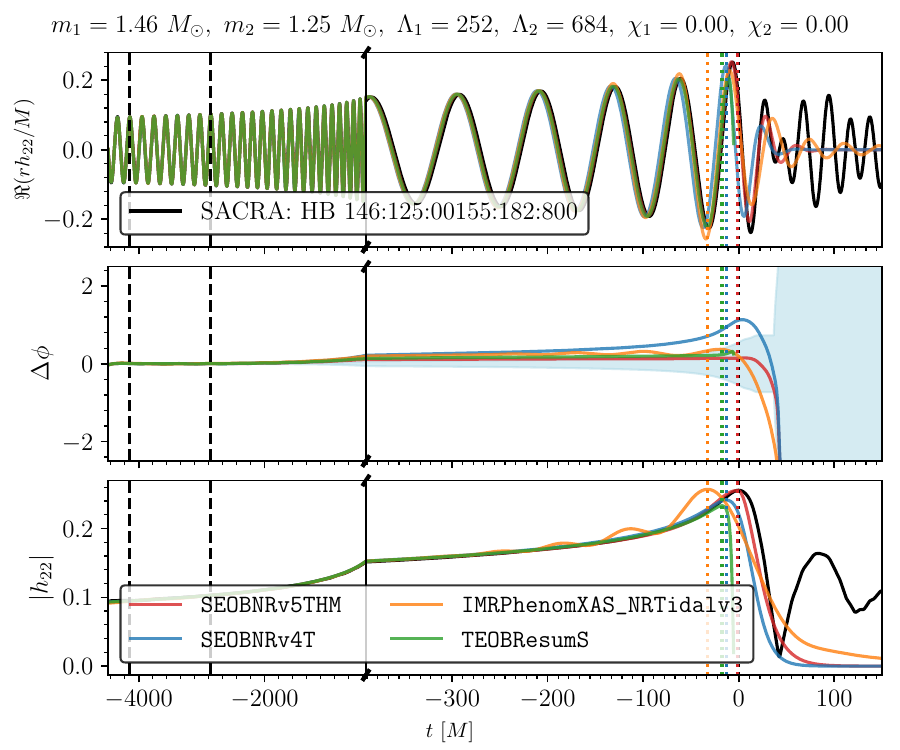} \\
   \includegraphics[width=0.5\linewidth]{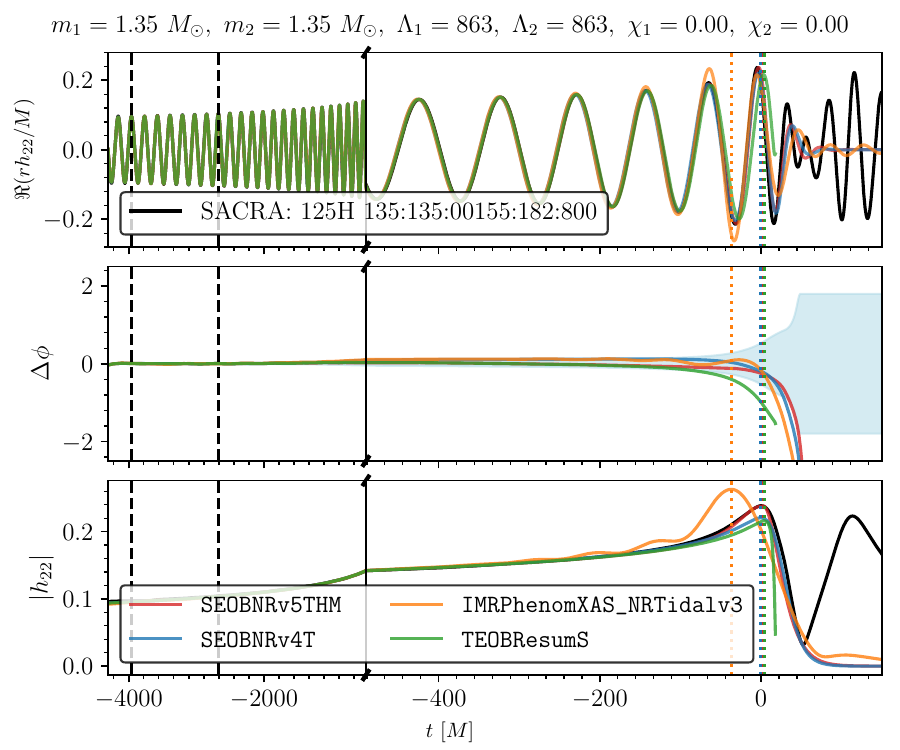}\hfill
   \includegraphics[width=0.5\linewidth]{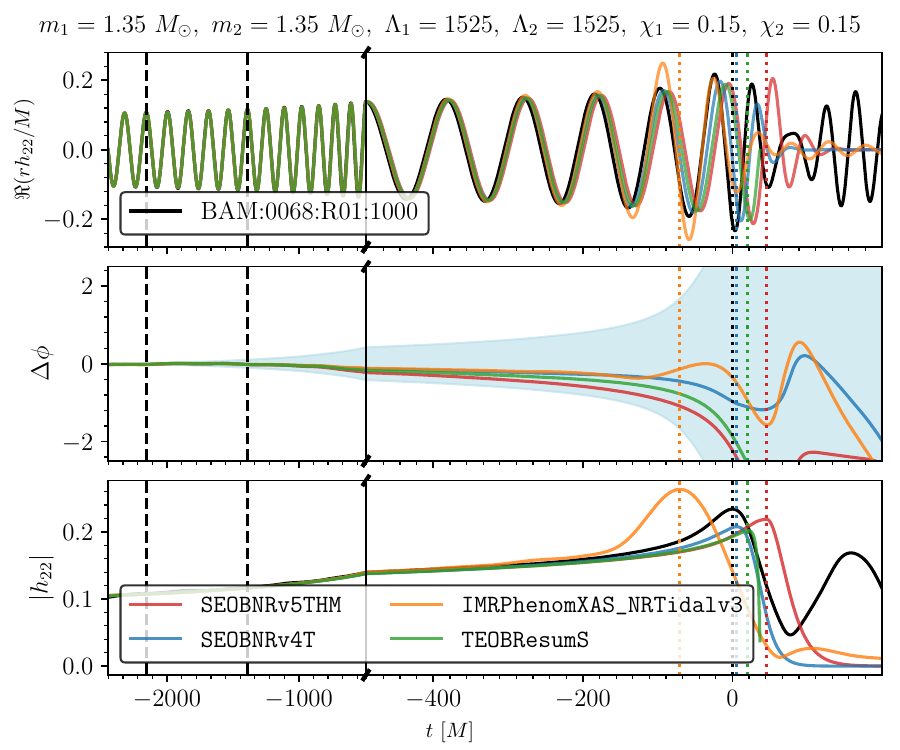} \\ 
   \includegraphics[width=0.5\linewidth]{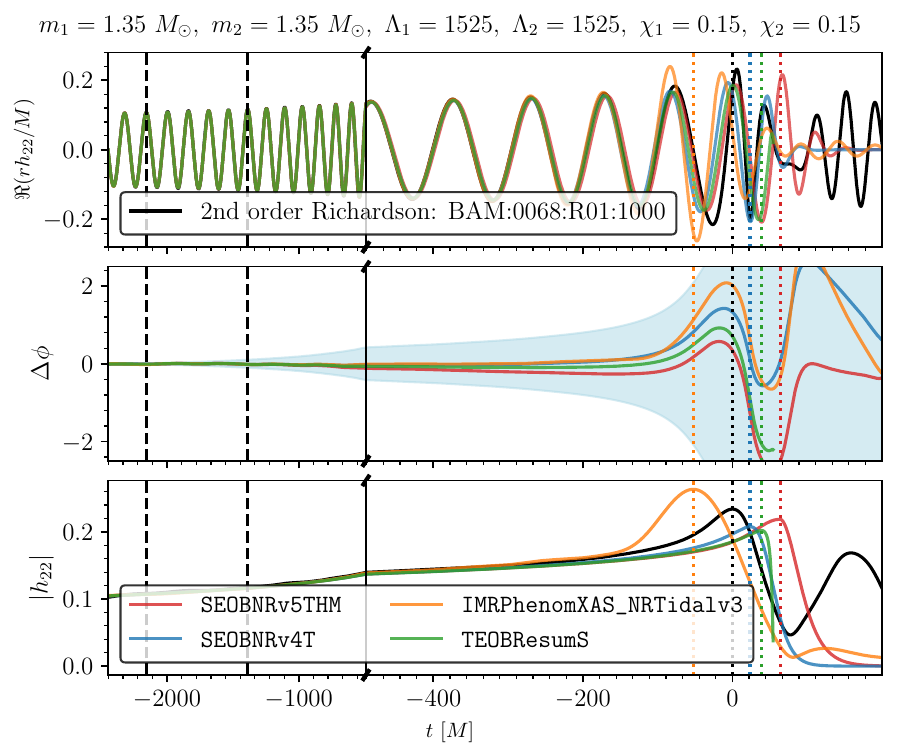}\hfill
   \includegraphics[width=0.5\linewidth]{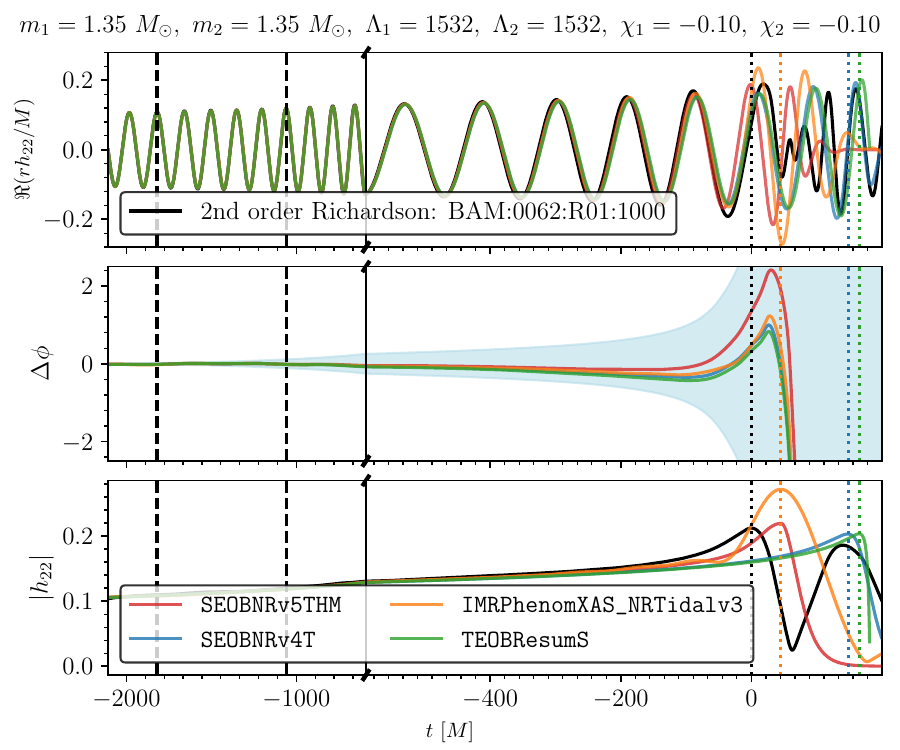} \\ 
   \caption{Time-domain comparisons for three SACRA and two BAM simulations against four approximants as different colors. For each NR simulation, we state the BNS parameters above the figure. For each figure, the upper panel shows the real part of the (2,2) mode, while the middle panel shows the phase difference between the waveform model and the NR waveform, and the lowest panel shows the amplitude of the (2,2) mode. We show the inspiral and merger with two different time scales to focus on the merger. We further indicate the merger times of the approximants with vertical dotted lines. We compare against the BAM:0068 simulation two times, once against the highest resolution simulation and once against the Richardson-extrapolated waveform at convergence order 2, which we indicate in the legend.}
   \label{fig:timedomaincomparisons}
\end{figure*}

\begin{figure*}[t]
    \centering
   \includegraphics[width=0.92\linewidth]{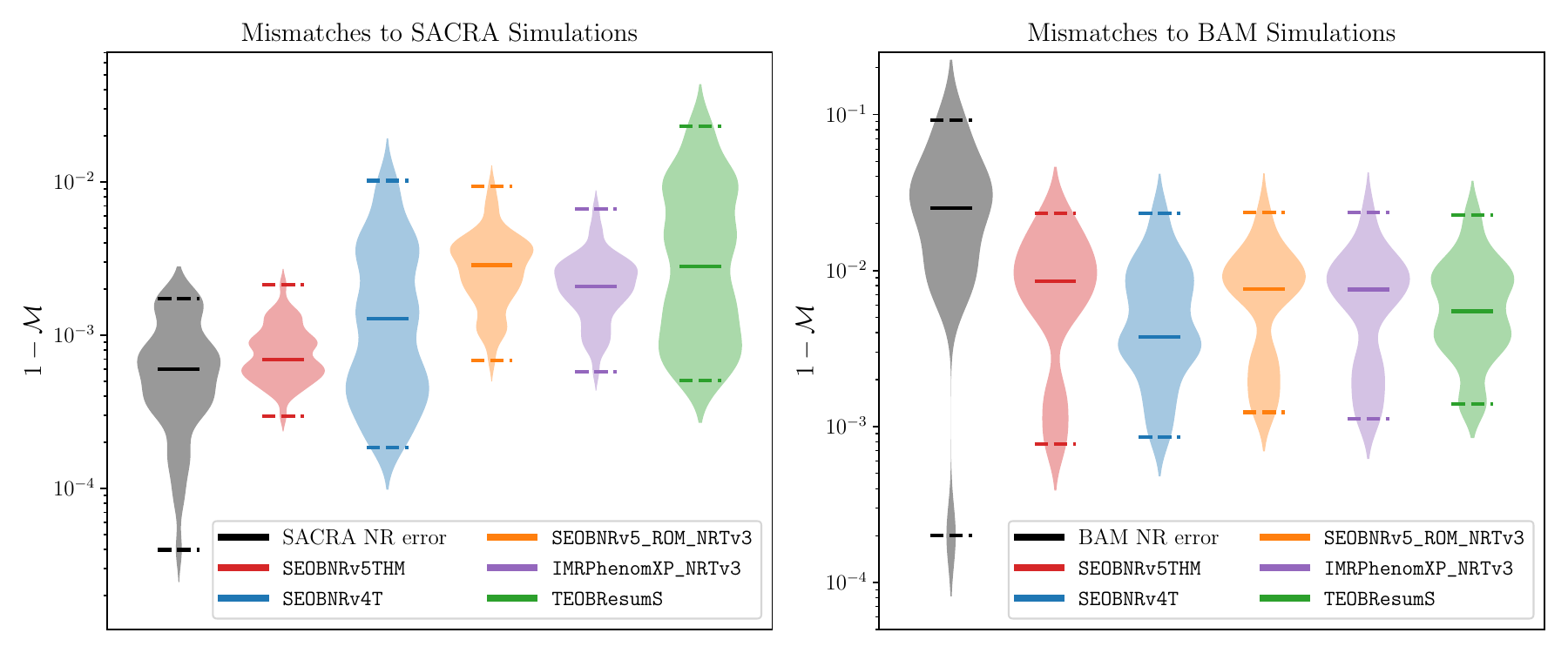}
   \caption{Mismatch comparison against SACRA waveforms on the left and BAM waveforms on the right, starting at 1.25 times the starting frequency of the simulation, and ending at the merger time for SACRA simulations, or the minimum of the merger times of the simulations used for the Richardson-extrapolation for BAM simulations. We use the advanced LIGO PSD and indicate with a solid line the median mismatch across all simulations and with dashed lines the minimum and maximum mismatch across all simulations. We highlight the different scales of the y-axis between the left and the right figure.}
   \label{fig:mm}
\end{figure*}

To compare the different approximants in time-domain to NR simulations, we use the same alignment procedure described in Eq.~\eqref{eq:alignment}. As the BAM waveforms are convergent at a clear convergence order, we Richardson-extrapolate the phase as in Ref.~\cite{Abac:NRT} to infinite resolution. To this end, we redefine the phase of the highest resolution NR waveform as
\begin{equation} \label{eq:Richardson}
    \phi_{\rm Richardson} = \frac{s\phi_{\rm high} - \phi_{\rm low}}{s - 1}\,,
\end{equation}
where
\begin{equation}
	s = \left(\frac{n_{\rm high}}{n_{\rm low}}\right)^k > 1\,,
\end{equation}
$k$ is the convergence order found through the standard procedure (see, e.g., Refs.~\cite{Gonzalez:2022mgo,Bernuzzi:2016pie}), $n$ are the number of grid points and $\phi$ is the phase of the highest and second highest (low) resolution waveforms.

As described in Sec.~\ref{sec:Calibration}, the merger time depends on the employed resolution of the NR simulation. When Richardson-extrapolating the waveform through the above equation, one therefore runs in the situation where one of the simulations has merged and the other is still in its pre-merger phase, which creates unphysical artefacts in the frequency and therefore phase. 

This effect is amplified when one instead tries to Richardson extrapolate the amplitude time-series, due to the peak at merger of the individual runs. If a lower-resolution run merges earlier than the highest-resolution run, a Richardson extrapolated amplitude will unphysically increase following the lower-resolution merger, which would distract from our results. As this scenario applies to nearly all of the NR simulations we study in this work, and as a correctly modeled phase is of more importance than a correctly modeled amplitude with regards to PE, we do not extrapolate the amplitude in our comparisons. We instead combine the highest-resolution amplitude with the Richardson extrapolated phases to arrive in our strain $h_{22}^{\rm NR}$ against which we then compare.

Similarly to what has been done for \nrtidalvthree\ in Ref.~\cite{Abac:NRT}, we also indicate the numerical uncertainty as shaded error bands in the phase difference $\Delta\phi = \phi_{22}^{\rm approx} - \phi_{22}^{\rm NR}$. In particular, as the SACRA simulations are not convergent at a clear convergence order, we calculate the phase difference between the highest and third-highest resolutions of a given system. The errors may be underestimated in this case. For NR simulations where we find a clear convergence order, we use the phase difference between the Richardson-extrapolated waveform and the highest resolution simulation as an error estimate.

The results of the time-domain comparison can be seen in Fig.~\ref{fig:timedomaincomparisons}. We want to highlight the effects of our pre-merger modeling as described in Sec.~\ref{subsec:Waveform} and of our calibration of the merger time as described in Sec.~\ref{sec:Calibration}. Although most of the approximants stay inside the error band of the phase difference $\Delta\phi$, it is evident that \vfivet\ consistently stays in and close to the NR phase all the way to merger for both SACRA and BAM.

\vfivet\ additionally predicts the merger time best for the SACRA simulations and models the amplitude correctly all the way up to merger. For BAM waveforms this is not necessarily the case, but we want to remind the reader that we do not Richardson-extrapolate the amplitude and that the merger time is therefore shown for the highest resolution BAM simulation, not the prediction for SACRA-level resolutions that we calibrated against, as described in Sec.~\ref{sec:Calibration}.

It is therefore to be  expected that our approach models the immediate amplitude of the highest resolution SACRA simulations correctly all the way up to merger, while predicting a later merger and hence a different amplitude evolution from the highest resolution BAM simulations. We hope that in the future, higher resolution BAM simulations will support our methodology of predicting the extrapolated merger time.

Note that for some simulations, the phase difference of all approximants may show jumps in Fig.~\ref{fig:timedomaincomparisons} during the post-merger. This is due to sudden changes in the NR-frequency $\omega_{22}^{\mathrm{NR}}$ soon after merger and prior to collapse of the remnant. This abrupt frequency modulation is associated to violent radial bounces of the remnant's core prior to collapse, see, e.g., Refs.~\cite{Bernuzzi:2015,Breschi:2022} and is not modeled by any approximant so far. As a consequence, all waveform models may show a discontinuous phase difference during the post-merger, if a simulation exhibits this feature.

It is furthermore interesting to note that the inclusion of the spin-shifted dynamical tides lowers the agreement with the highest resolution aligned-spin BAM simulations in comparison to \vfourt\ or \teob, but that it does perform comparable to these models when one looks at the Richardson-extrapolated waveform. As all approximants lie within the NR error band however, it is not yet possible to distinguish the accuracy of including the spin-shifted dynamical tides in BNS simulations (but see also the investigation in Ref.~\cite{Steinhoff:2021dsn}, including a comparison to high-accuracy simulations of NS-BH binaries). 

\subsection{Faithfulness against numerical relativity} \label{subsec:NR}

\begin{figure*}[t]
    \centering 
  \includegraphics[width=0.246\linewidth]{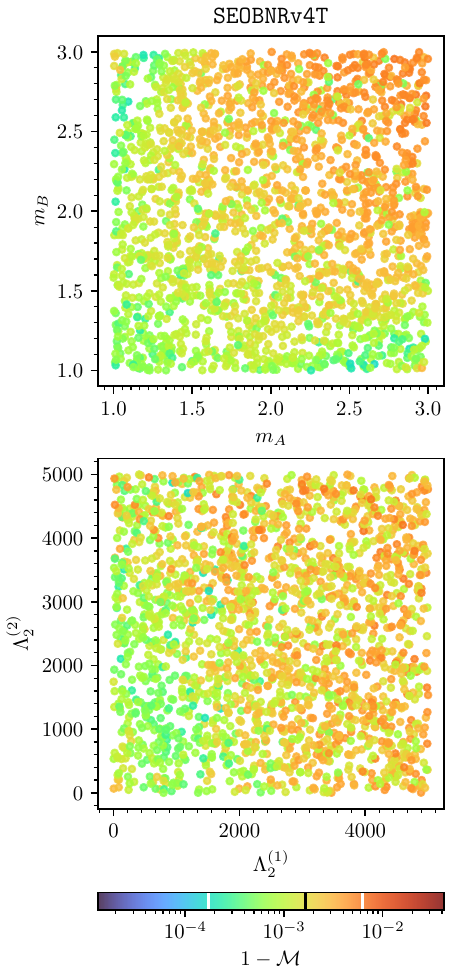}
  \includegraphics[width=0.246\linewidth]{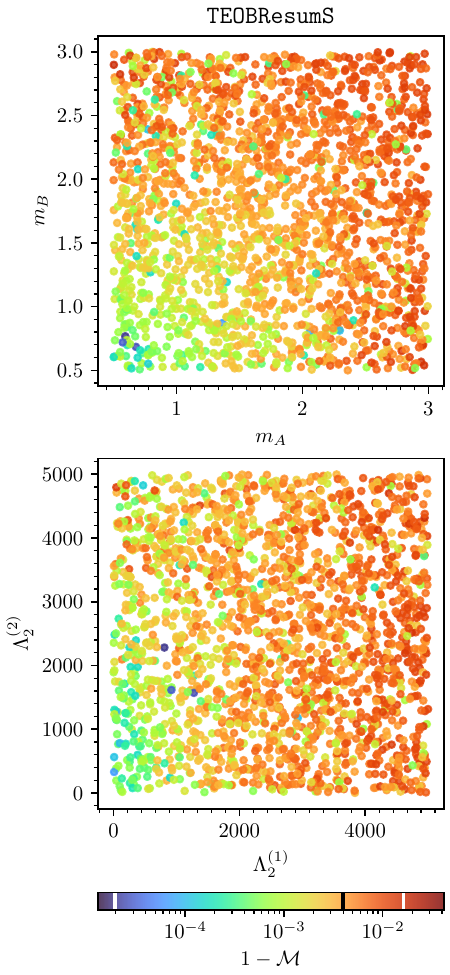}
  \includegraphics[width=0.246\linewidth]{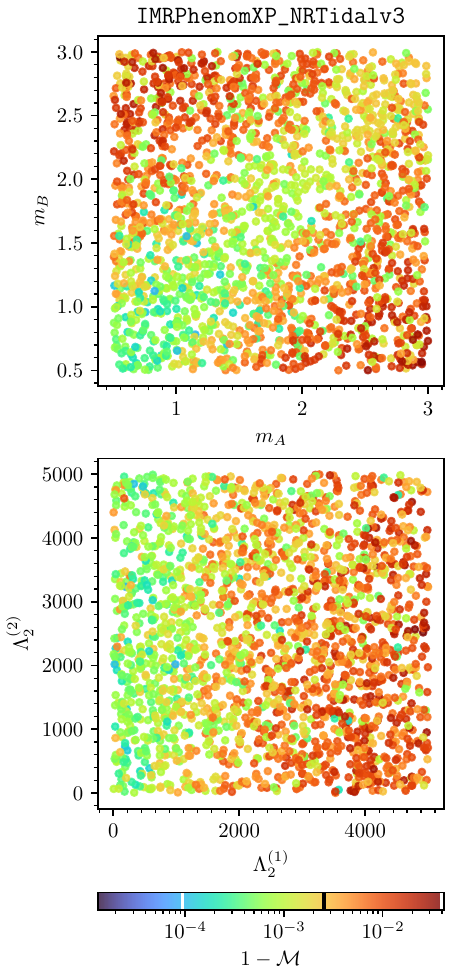}
  \includegraphics[width=0.246\linewidth]{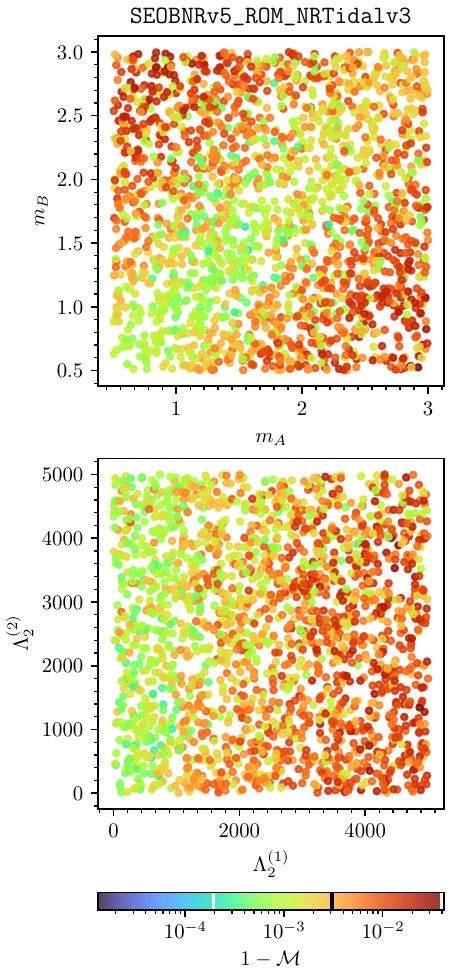}
\caption{Mismatch comparisons of \vfivet\ with other tidal waveform models for non-spinning configurations in each column. Each mismatch comparison contains 2000 random configurations with tidal deformability $\Lambda_2^{(1,2)} \in [0, 5000]$ and $m_{A,B} \in [0.5, 3]\ M_\odot$, except for \vfourt, where we set the mass-range to $m_{A,B} \in [1, 3]\ M_\odot$. For each subfigure we include a color bar indicating the values of the mismatches and state the name of the approximant in the title. We keep the minimum and maximum value in the color bar consistent across the different figures, but indicate the minimum (white), maximum (white), and median (black) mismatch against each approximant with solid vertical lines on the colorbar.}
\label{fig:mm_nospin}
   \end{figure*}

\begin{figure*}[t]
    \centering 
  \includegraphics[width=0.246\linewidth]{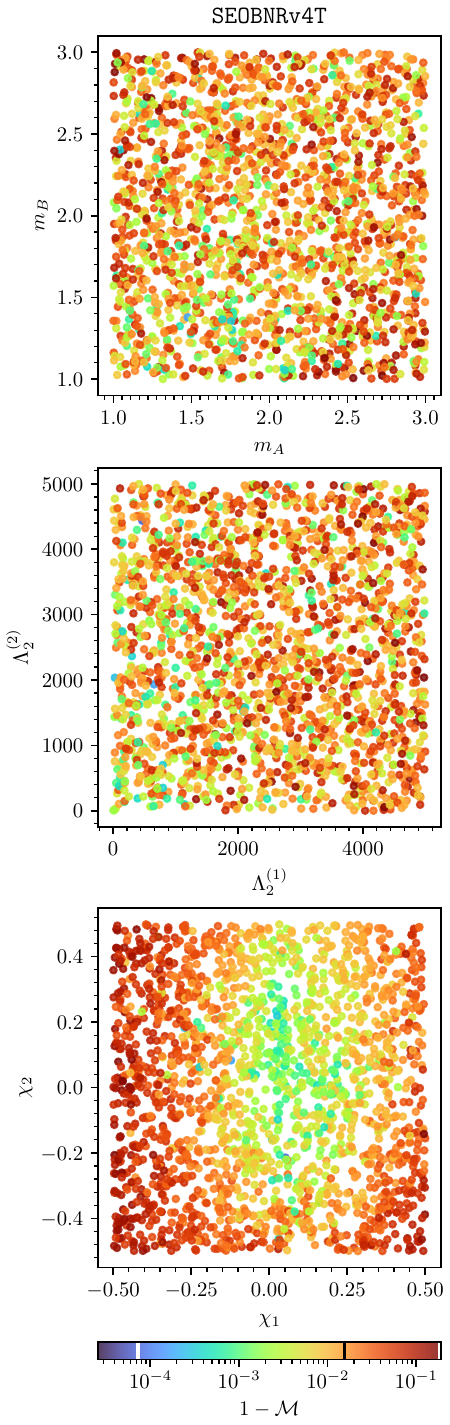}
  \includegraphics[width=0.246\linewidth]{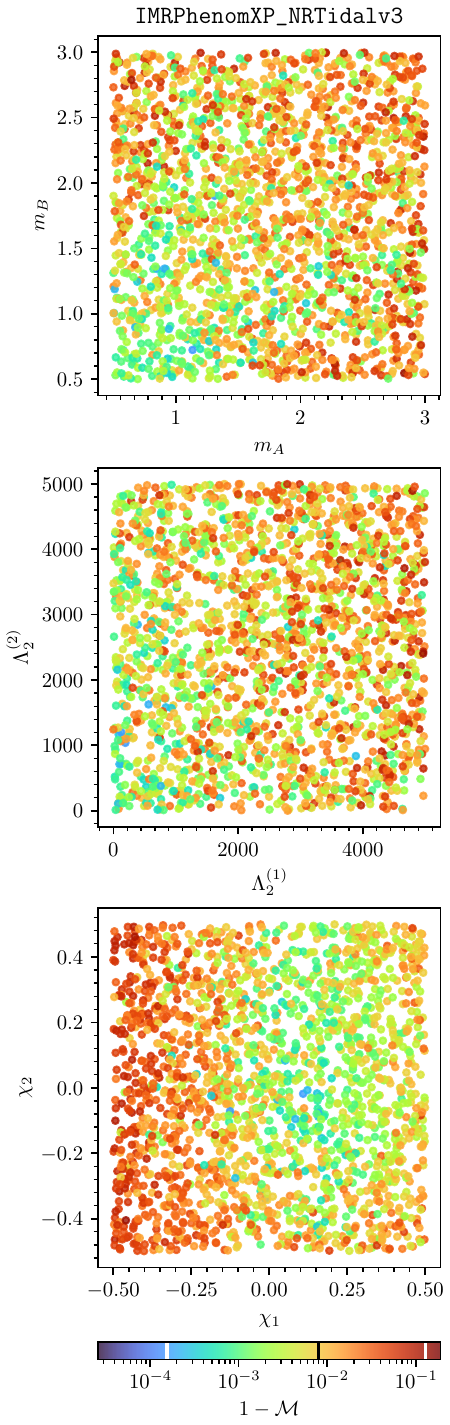}
  \includegraphics[width=0.246\linewidth]{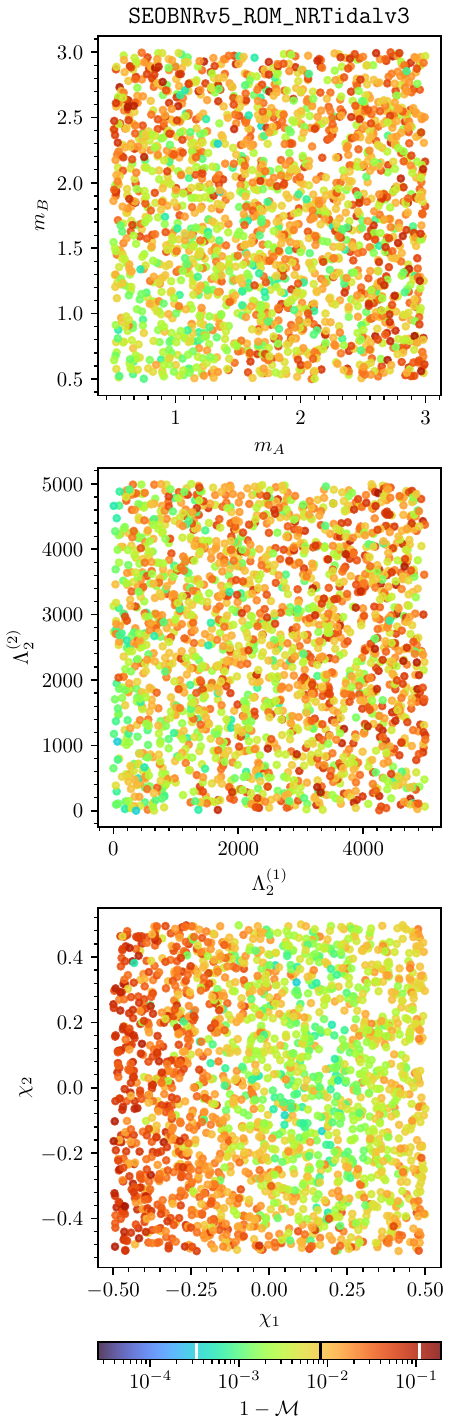}
  \includegraphics[width=0.246\linewidth]{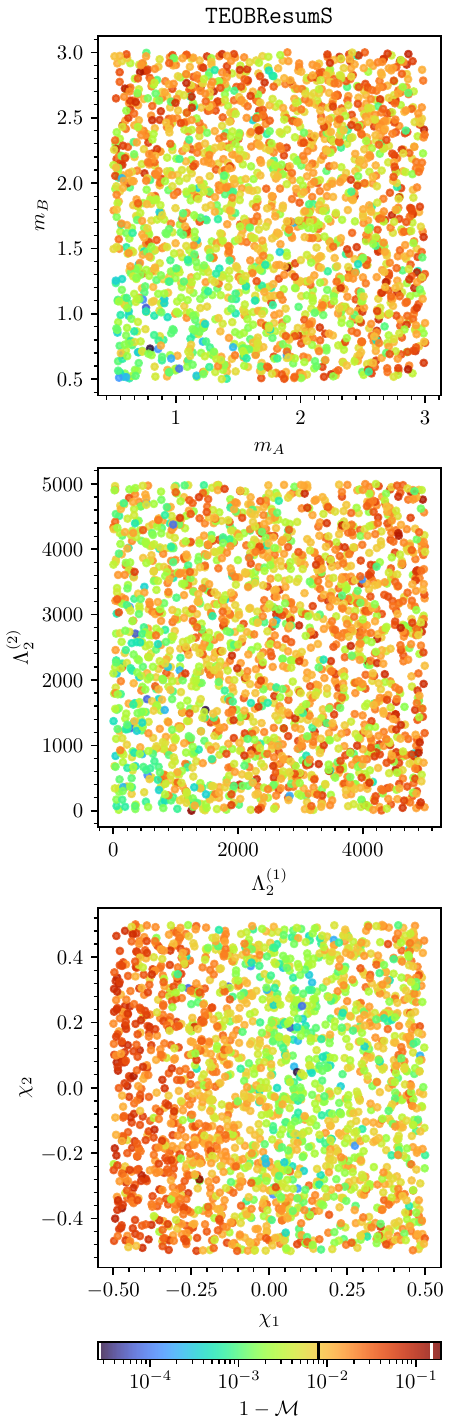}
\caption{Mismatch comparisons of \vfivet\ with other tidal waveform models for aligned-spin configurations in each column. Each subfigure has 2000 random configurations of mass $m_{A,B} = [0.5, 3]$, tidal deformability $\Lambda_2^{(1,2)} = [0, 5000]$ and spin $\chi_{1,2} = [-0.5,0.5]$, except for \vfourt, where the mass-range is $m_{A,B} = [1, 3]$. For each subfigure we include a color bar indicating the values of the mismatches and state the name of the approximant in the title. We keep the minimum and maximum value in the color bar consistent across the different figures, but indicate the minimum (white), maximum (white), and median (black) mismatch against each approximant with solid vertical lines on the colorbar.}
\label{fig:mm_spin}
\end{figure*}

We want to compare all available BNS approximants against all the NR simulations used in this work by computing mismatches between the approximants and simulations. The mismatch (or unfaithfulness) between two waveforms ${h}_1(t)$ and ${h}_2(t)$ is given by
\begin{equation}\label{eq:mismatch}
    1 - \mathcal{M} = 1 - \max_{(\phi_c, t_c)}\frac{(h_1(\phi_c, t_c)|h_2)}{\sqrt{(h_1|h_1)(h_2|h_2)}},
\end{equation}
where the overlap is defined in frequency-domain as
\begin{equation}\label{eq:overlap}
    (h_1|h_2) = 4 \,{\rm Re} \int_{f_\text{min}}^{f_{\text{max}}}\frac{\tilde{h}_1^{*}(f)\tilde{h}_2(f)}{S_n(f)}df.
\end{equation}

The maximisation of the overlap over some arbitrary phase $\phi_c$ and time shift $t_c$ in the time-domain corresponds to a phase and frequency shift in frequency-domain and ensures the alignment between the two waveforms similar to Eq.~\eqref{eq:alignment}. In line with the procedure for \vfivehm\ in Ref.~\cite{Pompiliv5}, we use the design zero-detuned high-power noise power-spectral-density (PSD) of Advanced LIGO from Ref.~\cite{Barsotti:2018} for our mismatch computations.

We taper the beginning of time-domain waveforms with a Planck window, and to ensure no mismatch contributions from the tapering we choose the lower bound of the mismatch integral as $f_{\text{min}} = 1.25\,f_{\text{start}}$, with $f_{\text{start}}$ the point of the NR waveform where the junk radiation has decayed.

To estimate the NR error, we compute mismatches between the waveforms that were also used to create the error bands in the previous subsection, i.e. the third-highest resolution SACRA waveforms or the highest resolution waveforms for BAM, where we compare the approximants against the Richardson-extrapolated waveforms. 

For SACRA we then set the upper bound of the mismatch integral as $f_{\text{max}} = f_{\text{mrg}}$ of the highest resolution simulation. For BAM on the other hand, we choose the frequency of the Richardson-extrapolated waveform that corresponds to the minimum merger time of the two simulations used in the Richardson-extrapolation
\begin{equation}\label{eq:Richardson_f}
    f_{\text{max}} = f_{\text{Richardson}}\left(\min (t_{\rm peak}^{\rm high},\,t_{\rm peak}^{\rm low})\right).
\end{equation}
We employ this procedure to reduce the NR error mismatches in comparison to BAM, which are otherwise larger due to artefacts in the Richardson-extrapolated waveforms beyond this time.

As the \teob\ waveform terminates at merger, see Ref.~\cite{Gamba:2023mww}, we taper the waveform with a similar approach as in \vfourt\ prior to Fourier transforming. We choose a sampling rate of $2^{13} = 8192$ Hz for all of the approximants, only compare the $(2,2)$ mode of all waveforms, and show our results as violin plots of the smoothed mismatch histograms in Fig.~\ref{fig:mm}.

We note that when looking at the 45 SACRA simulations, \vfivet\ has the lowest maximum and median mismatch of all BNS approximants and remains comparable to the SACRA NR error. This shows the overall accuracy of the model due to the calibration and pre-merger modeling. When comparing against BAM, the model performs slightly worse, but comparable to the other approximants. We however want to raise awareness to the higher NR error mismatches for these simulations. It is therefore hard to decide on a best approximant for the BAM simulations, as all approximants lie within the NR error.

The low outlier mismatch of order $10^{-4}$ corresponds to BAM:0001, which has a very low mismatch but also possesses the lowest tidal deformabilities of all the BNS simulations at hand. Note however that the highest-resolution run of BAM:0001 is rather low compared to the resolutions of the other BAM simulations, as can be seen in Fig.~\ref{fig:BAM_t_peak}. The NR error computed with the Richardson-extrapolation might therefore be underestimated in this particular case.

\subsection{Mismatch comparison to other tidal waveform approximants} \label{subsec:Mismatches}

Finally, we want to compare our model with other tidal models for non-spinning and aligned-spin configurations across the relevant BNS parameter space. For BNS systems we therefore compute the mismatch between \vfivet\ and \teob, \vfourt, as well as \seobnrvfiveromnrtidalvthree\ and \imrphenomxpnrtidalthree. 

All mismatches computed in this section are performed as in Eq.~\eqref{eq:mismatch} for the $(2,2)$ mode of the approximants between $f_{\rm min} = 20\, {\rm Hz}$, and $f_{\rm max} = 2048\, {\rm Hz}$. For computational efficiency we employ the \vfourt-surrogate from Ref.~\cite{Lackey:2018zvw} instead of \vfourt\ directly. We call \vfivet\ consistently in frequency-domain or time-domain dependent on the native domain of the approximant it is compared to, to avoid artefacts from applying different Fourier transform prescriptions to the individual waveforms. When comparing against the time-domain waveform \teob, we use sampling rates of $2^{13} = 8192 \, {\rm Hz}$, while we use $\Delta f=1/1024 \, {\rm Hz}$ when comparing to the frequency-domain waveforms \vfourt-surrogate, \nrtidalvthree\ and \vfivehm, which we call in frequency-domain.

In the non-spinning case, the mismatches are computed for 2000 random configurations of masses $m_{A,B} \in [0.5, 3]\ M_{\odot}$ and tidal deformabilities $\Lambda_2^{(1,2)} \in [0, 5000]$, for comparisons done with \teob\ and \nrtidalvthree. We reduce the mass-range to $m_{A,B} \in [1, 3]\ M_{\odot}$ for the \vfourt-surrogate, which is only available up to mass ratio $q \leq 3$. We find that the mismatch increases with higher total mass, as the effect of the differing pre-merger prescriptions of the waveform approximants reaches more of the sensitive detector band. The reduction of the mass-range for lower mass systems should therefore not affect our results for \vfourt\ drastically in comparison to the other waveform approximants.

The changes that we have introduced into \vfivet\ in comparison to its baseline model \vfivehm, namely the omission of the phase NQCs and our novel pre-merger modeling approach, additionally yield differences in the waveform compared to \vfivehm\ even when the tidal deformabilities are set to zero. We also check that these changes have a negligible impact on PE by computing the mismatches of spin-aligned BBH systems between the two approximants.

For the BBH model comparison with \vfivehm\ we reduce the mass-range to $m_{A,B} \in [1, 3]\ M_{\odot}$, set $\Lambda_2^{(1,2)}=0$ for \vfivet, and consider an additional spin range $\chi_{1,2} \in [-0.5,0.5]$. \vfivehm\ models the ringdown of the resulting BH and requires the QNMs to be resolved, which means that we need a high sampling frequency of the low total-mass BBH for this waveform model. For consistency, we thus call both \texttt{SEOBNRv5} approximants in frequency-domain up to a maximum frequency of $2^{15} = 32\,768 \, {\rm Hz}$ (which translates into a high sampling frequency in time-domain) when comparing them. The mismatches to \vfivehm\ stay below the $1.2\times 10^{-3}$ limit across the described BBH parameter space and are therefore not shown in this work. We find that the uncertainty in BNS waveform modeling introduced by tidal effects dominates the mismatch and therefore PE, and our changes to the BBH waveform will have a negligible systematic effect in the study of BNS signals.

\begin{figure*}[t]
    \centering
    \includegraphics[width = 0.99\linewidth]{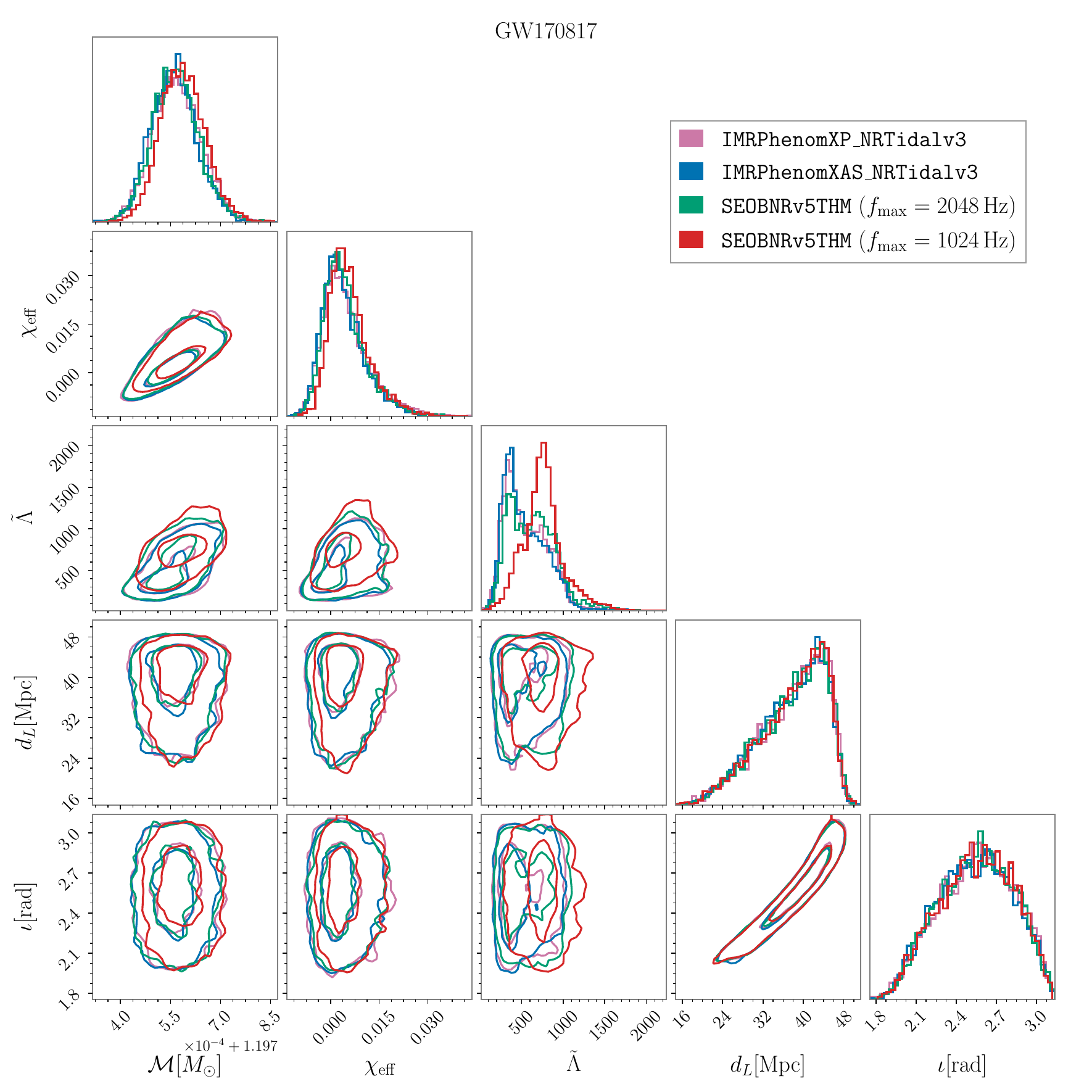}
    \caption{The marginalized 1D and 2D posterior probability distributions for selected parameters of GW170817, obtained with \imrphenomxpnrtidalthree\ (blue), \xasnrtidalvthree\ (orange), and \vfivet\ (green/red). We run \vfivet\ twice with different high-frequency cutoffs, at either 2048 Hz, as used for the \nrtidalvthree-approximants (green), and at 1024 Hz in line with the analysis in Ref.~\cite{Gamba:2021prd}. The parameters shown here are the detector-frame chirp mass of the system $\mathcal{M}$ in solar masses, the effective spin $\chi_{\rm eff}$, the binary tidal deformability $\tilde{\Lambda}$, the luminosity distance $d_L$ in Mpc, and the inclination angle $\iota_0$. The 68\% and 90\% confidence intervals are indicated by contours for the 2D posterior plots.}
    \label{fig:PE_GW170817}
\end{figure*}

The resulting mismatches for the BNS approximants can be seen in Fig.~\ref{fig:mm_nospin}. Note that while we assume the mass ordering of the primary and secondary $m_1 \geq m_2$, we will use randomly chosen $m_A$ and $m_B$ in the top row of this figure to ignore this mass ordering. Otherwise, only half of the figure would be filled with data points. In the lower rows however, we show the tidal deformability $\Lambda_2^{(1)}$ ($\Lambda_2^{(2)}$)
associated to the primary (secondary). This results in the individual subfigures possessing different symmetries.

Our findings are in-line with the results of a similar study performed for \nrtidalvthree\ in Ref.~\cite{Abac:NRT}, and our mismatches with \vfourt\ and \teob\ are lower than the reported mismatches of \nrtidalvthree\ with \vfourt\ and \teob. This is in line with our finding that the highest mismatches are in comparison to the \nrtidalvthree\ models, which goes up to a few $10^{-2}$. Note also that for near equal-mass systems, the \nrtidalvthree\ models have lower mismatches to \vfivet\ than the other EOB BNS models have to \vfivet. For unequal-mass systems however, the mismatches between \nrtidalvthree\ models and \vfivet\ increase quite strongly. This mismatch dependence on mass ratio for \texttt{NRTidalv2}\ and \nrtidalvthree\ has been previously reported in Ref.~\cite{Dietrich:2019kaq,Abac:NRT} and appears to be a systematic effect for this class of approximants. It might be insightful to study its origin in the future. A possible hypothesis to be tested would be, for example, that this is caused by one of the analytical approximations on which the \nrtidal-family is built being more valid for equal mass binaries than for unequal mass ones. Namely, the approximation that the GW phase can be linearly decomposed into different contributions, one of which being the tidal contribution~\cite{Dietrich:2017aum}.

The mismatches between \vfivet\ and the other EOB approximants \vfourt\ and \teob\ are below or around the $10^{-2}$ level across the non-spinning BNS parameter space. This serves as a consistency check that there are no pathologies in the model and that \vfivet\ stays in line with past results. The largest mismatches with the other EOB BNS models are found for high primary tidal deformability $\Lambda_2^{(1)}$ and high total mass $M$. Another noteworthy result is that the change of BBH baseline between \seobnrvfiveromnrtidalvthree\ and \imrphenomxpnrtidalthree\ only plays a minor role in the mismatches of non-spinning BNS systems.

In addition to the 2000 random configurations of the masses and tidal deformabilities that we generated for the non-spinning case, we also generate 2000 random aligned-spin BNS waveforms $\chi_{1,2} = [-0.5,0.5]$ for the comparison of \vfivet\ against \vfourt, \seobnrvfiveromnrtidalvthree, \imrphenomxpnrtidalthree\ and \teob, which can be seen in Fig.~\ref{fig:mm_spin}. For context, the maximum dimensionless spin an NS can have before it breaks up is about 0.67 \cite{Lo:11}, or about a spin rate of 1000 Hz (depending on the equation of state). 
The fastest-spinning pulsar known, J1748-2446ad, has a spin rate of 716 Hz~\cite{Hessels:06}, which translates into a dimensionless spin of about 0.4~\cite{Nielsen:2016}. It furthermore resides in the globular cluster Terzan 5 and has a light companion, which makes it a candidate for the production of GW sources due to dynamical capture in its future~\cite{Rodriguez:15,Yu:2024}.

\begin{figure*}[t]
    \centering
    \includegraphics[width = 0.95\linewidth]{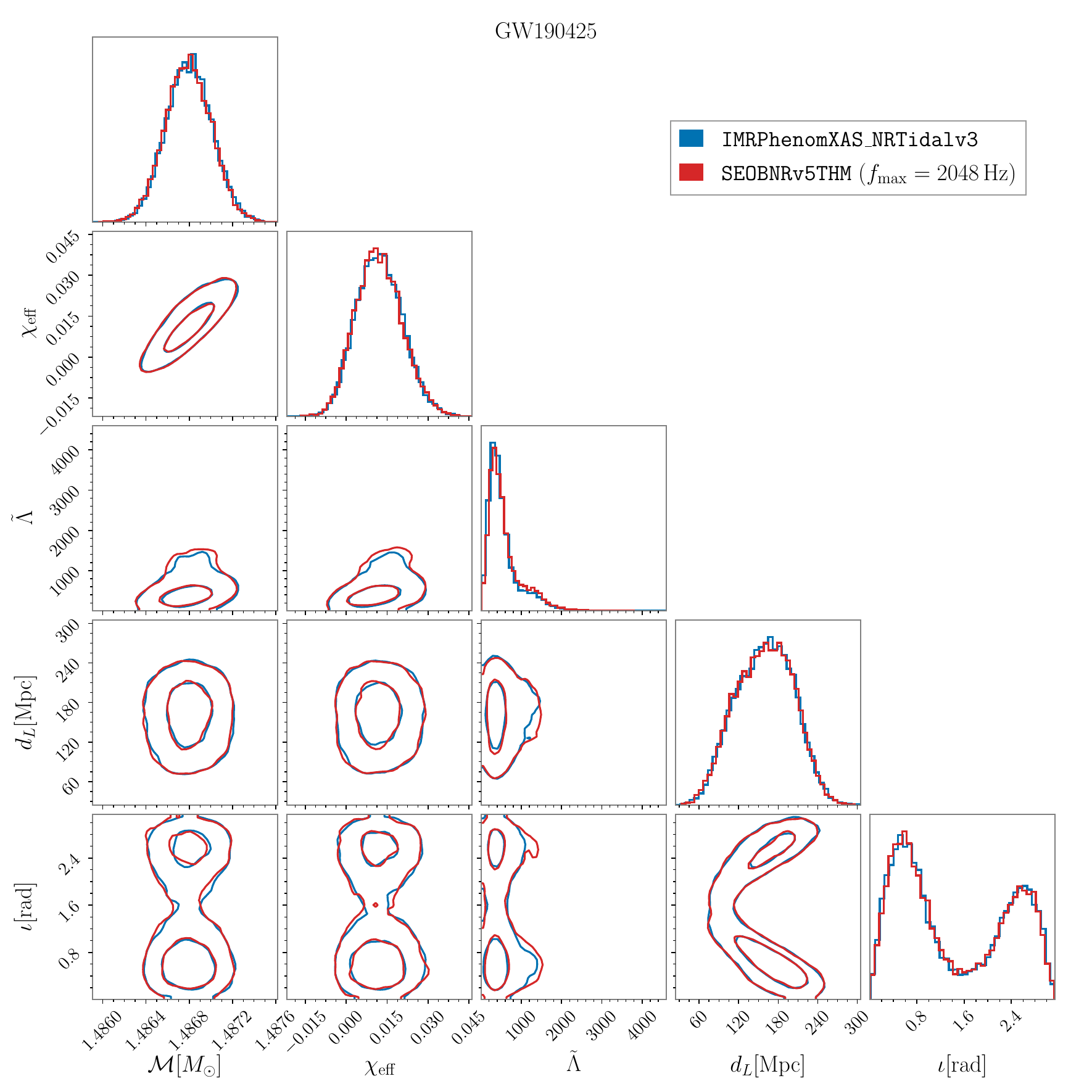}
    \caption{The marginalized 1D and 2D posterior probability distributions for selected parameters of GW190425, obtained with \xasnrtidalvthree\ (blue), and \vfivet\ (red) up to a high-frequency cutoff of 2048 Hz. The parameters shown here are the detector-frame chirp mass of the system $\mathcal{M}$ in solar masses, the effective spin $\chi_{\rm eff}$, the binary tidal deformability $\tilde{\Lambda}$, the luminosity distance $d_L$ in Mpc, and the inclination angle $\iota$. The 68\% and 90\% confidence intervals are indicated by contours for the 2D posterior plots.} 
    \label{fig:PE_GW190425}
\end{figure*}

The mismatches here are much higher than for non-spinning systems and are mainly driven by highly spinning BNS configurations, with the highest mismatches being found for non-zero primary's spin $|\chi_1| \gtrsim 0.2$ and in particular for anti-aligned BNS systems $\chi_1 \lesssim -0.25$. The mismatches are the highest with \vfourt, which does not include the effect of spin-induced multipole moments above the LO quadrupole, in contrast to all other approximants studied in this work. For highly spinning systems the mismatches can approach and exceed the $10^{-1}$ level, which is due to the spin-shifted dynamical tides~\cite{Steinhoff:2021dsn} that have not been included in another waveform model that has been reviewed by the LVK Collaboration so far.

Since we terminate anti-aligned BNS systems with large spin-magnitudes prior to the development of eccentricity as described in Sec.~\ref{subsec:Hamiltonian}, these waveforms are much shorter than the waveforms from approximants that do not include spin-shifted dynamical tides. Note that there is evidence from NR that such systems indeed have a much shorter inspiral-phase~\cite{Kuan:2024}, although further studies are needed. For now, this early merger that we find for \vfivet\ should explain the high mismatches up to the order of $10^{-1}$ for binary systems containing highly spinning, anti-aligned NSs in Fig.~\ref{fig:mm_spin}.

We leave the correct modeling of post-resonance BNS waveforms as future work and note that the approximant should not be blindly trusted for NSs with large negative spins $|\chi_i| \lesssim 0.3$. We have still included the spin-shift effect in our model, as it is a theoretical prediction valid for low spins and serves as a good starting point to incorporate future analytical developments.

\section{Parameter Estimation} \label{sec:PE}

\begin{table}
    \centering
    \resizebox{\columnwidth}{!}{%
    \begin{tabular}{l c c c c c c }
     \hline
     \hline
      \makecell[cc]{Event} & \makecell[cc]{Model} & \makecell[cc]{Data \\ settings} &\multicolumn{2}{c}{\makecell[cc]{Sampler \\ settings}} & \makecell[cc]{Computing \\ resources} & Runtime  \\
     \hline
     & & \makecell[cc]{$f_\text{max}$ \\ $\left[ \text{Hz} \right]$}
     & nact & nlive & nodes$\times$cores &   \\
     \hline
       \multirow{3.3}{*}{ \makecell[ccc]{  GW170817	}
       } 
       &   \makecell[cc]{\fontsize{7.5pt}{7pt}\selectfont  \texttt{IMRPhenomXAS} \\[-0.1cm] \fontsize{7.5pt}{7pt}\selectfont  \texttt{\_NRTidalv3}} & 2048 & 65 & 1024 & $12\times 32$ & 15h \\[0.08cm]
        & \makecell[cc]{\fontsize{7.5pt}{7pt}\selectfont  \texttt{SEOBNRv5THM}}   & 2048 & 65 & 1024 & $16\times 32$ & 1d 6h  \\[0.08cm]
        & \makecell[cc]{\fontsize{7.5pt}{7pt}\selectfont  \texttt{SEOBNRv5THM}}   & 1024 & 20 & 1024 & $12\times 32$ & 18h \\
    \hline
    \multirow{2.3}{*}{ \makecell[ccc]{  GW190425	}
    }  &   \makecell[cc]{\fontsize{7.5pt}{7pt}\selectfont  \texttt{IMRPhenomXAS} \\[-0.1cm] \fontsize{7.5pt}{7pt}\selectfont  \texttt{\_NRTidalv3}} & 2048 & 65 & 2048 & $16\times 32$ & 1d 3h \\[0.08cm]
     & \makecell[cc]{\fontsize{7.5pt}{7pt}\selectfont  \texttt{SEOBNRv5THM}}   & 2048 & 65 & 2048 & $16\times 32$ & 1d 20h  \\
    \hline \hline
    \end{tabular}
    }
    \caption{
    Settings and evaluation times for the different PE runs on two real GW events GW170817 and GW190425 performed in this work. We employ the {\tt parallel} {\tt Bilby}~\cite{Smith:2019ucc} sampler with the waveform models \texttt{SEOBNRv5THM} and \xasnrtidalvthree.
    We list the maximum frequency ($f_\mathrm{max}$) starting from $f_\mathrm{min}=20$ Hz in the data settings, while the number of live points ($\text{nlive}$), and the number of autocorrelation times ($\text{nact}$) are specified in the sampler settings.
    We use a duration of $ 128\,$s for all the runs in this table.
    The time reported is walltime, while the total computational cost in CPU hours can be obtained by multiplying this time by the reported number of CPU cores employed. In general, we find that an analysis using \texttt{SEOBNRv5THM} takes on the order of twice as long as an analysis using \xasnrtidalvthree.
    }
    \label{tab:peruntime}
    \end{table}

As a last validation to test the applicability of the \vfivet\ model for data-analysis, we reanalyze the two BNS detections GW170817~\cite{LIGOScientific:2017vwq} and GW190425~\cite{LIGOScientific:2020aai} of the LVK Collaboration.
For this purpose, we utilize \texttt{parallel\ bilby}~\cite{Smith:2019ucc}, which performs GW parameter estimation using a parallelized nested sampler named \texttt{dynesty}~\cite{Speagle:2019ivv}. 
In \texttt{parallel\ bilby}, the Bayes' theorem is employed
\begin{equation}
    P(\bm{\theta}|d, \mathcal{H}) = \frac{\mathcal{L}(d|\bm{\theta}, \mathcal{H})p(\bm{\theta}|\mathcal{H})}{E(d|\mathcal{H})},
\end{equation}
where $P(\bm{\theta}|d, \mathcal{H})$ is the posterior probability distribution of the waveform parameters $\bf \theta$ given some data $d$, which in our case consists of the observed strain-data, and hypothesis $\mathcal{H}$. $p(\bm{\theta}|\mathcal{H})$ is the prior probability distribution, and $E(d|\mathcal{H})$ is the evidence, serving as normalisation constant to $P(\bm{\theta}|d, \mathcal{H})$. Meanwhile, $\mathcal{L}(d|\bm{\theta}, \mathcal{H})$ is the likelihood of obtaining the data $d$ containing a waveform with parameters $\bm{\theta}$ given the hypothesis $\mathcal{H}$, derived from a Gaussian noise model with power spectral density extracted from the data around the GW event.  For further details, we refer to Refs.~\cite{ Romero-Shaw:2020owr, Ashton:2021cub}.
In our study, we use GW170817 and GW190425 using the Open Data from Ref.~\cite{Abbott_2023}, and analyze these events with \vfivet, \imrphenomxpnrtidalthree\ (where we use the posteriors from the study performed in Ref.~\cite{Abac:NRT}), and \xasnrtidalvthree, and also compare their results with each other. For the analysis of both events, the LVK Collaboration employed a low-spin prior $|\chi|\leq 0.05$ and a high-spin prior analysis $|\chi|\leq 0.89$~\cite{LIGOScientific:2017vwq,LIGOScientific:2020aai}. For this work we have opted to only repeat the low-spin prior analysis. We remind our finding in Sec.~\ref{subsec:Mismatches} that the tidal approximants should not be blindly trusted for highly spinning NSs $|\chi|\gtrsim 0.3$. We follow the standard LVK Collaboration analyses~\cite{LIGOScientific:2018hze, LIGOScientific:2017vwq, Romero-Shaw:2020owr, Dietrich:2020efo, LIGOScientific:2020aai, Ashton:2021cub}.

The data settings, sampler settings, and computing resources for the individual analyses performed in this work are listed in Table~\ref{tab:peruntime}. We highlight that when studying the total cost in terms of the CPU hours, \vfivet\ appears to be $\mathcal{O}$(twice) as costly as the same analysis when employing \xasnrtidalvthree. It appears that the reason for this is the order 1~s overhead per likelihood evaluation in \texttt{parallel\ bilby}\footnote{We calculate the overhead by accessing the number of likelihood evaluations $N$ and the total sampling time $T$ from the result file of the respective \texttt{parallel\ bilby} runs. The time for one likelihood evaluation is then estimated as $T n_\mathrm{nodes} n_\mathrm{cores}/N$ with $n_\mathrm{nodes}$ ($n_\mathrm{cores}$) the number of employed nodes (cores). The likelihood evaluation time is on average $2.6$~s for \vfivet\ and $1.4$~s for \xasnrtidalvthree\ across both events studied in this work at $f_\mathrm{max}=2048$~Hz.}, which is shared between \vfivet\ and \xasnrtidalvthree. The effect of the much higher waveform generation speed of \xasnrtidalvthree\ is therefore reduced during PE.

We show the results of the inferred marginalized 1D and 2D posterior probability distribution of a selection of source parameters for GW170817 in Fig.~\ref{fig:PE_GW170817}, for a low-spin prior with $\chi \le 0.05$.  We show the detector-frame chirp mass of the system $\mathcal{M}$ in solar masses, the effective spin $\chi_{\rm eff}$ from Eq.~\eqref{eq:spins}, the luminosity distance $d_L$ in Mpc, and the inclination angle $\iota$. We also include here the dimensionless tidal deformability defined in terms of the individual masses and tidal deformabilities of the stars, see Ref.~\cite{Dietrich:2020efo}, which is related to $\kappa_2^T$
\begin{equation}
    \tilde{\Lambda} = \frac{16}{13}\frac{(m_1+ 12m_2)m_1^4\Lambda_2^{(1)} + (m_2 + 12m_1)m_2^4\Lambda_2^{(2)}}{M^5}.
\end{equation}
We further fix the sky position as recovered from the electromagnetic counterpart for GW170817.

From Fig.~\ref{fig:PE_GW170817}, we note that the performance of \vfivet\ in terms of the inferred parameters is consistent with the results of  \imrphenompvtwonrtidaltwo~\cite{LIGOScientific:2018hze} when analysing the signal up to 2048 Hz.

The main difference is the slightly larger secondary peak of \vfivet\ for $\tilde{\Lambda}$ compared to the other models. Note that all the posterior weight goes towards this secondary peak when analysing the event only up to 1024 Hz.

It has been shown in Refs.~\cite{Gamba:2021prd,Dai:2018,Narikawa:2019} that the first peak might be related to the glitch mitigation in Livingston. As discussed in Refs.~\cite{Gamba:2021prd}, above 1 kHz only a signal-to-noise of $\lesssim 1$ should have been deposited for GW170817-like events (which
roughly corresponds to 3\% of the total SNR), further hinting at the origin of the first peak being related to data quality issues. When only performing PE up to a high-frequency cutoff of 1 kHz, our results are furthermore fully consistent with the results for \teob\ in Ref.~\cite{Gamba:2021prd}.

The results for the marginalized posterior distributions for GW190425 can be found in Fig.~\ref{fig:PE_GW190425}, where all results are fully consistent between \xasnrtidalvthree\ and \vfivet.

\section{Conclusion} \label{sec:Conclusion}

In this work, we introduce the new waveform approximant \vfivet\ for BNS systems. This model represents a significant advancement over the previous model \vfourt~\cite{Steinhoff:2016rfi,Lackey:2018zvw}. The key improvements include
\begin{enumerate}[label=(\roman*)]
    \itemsep-3pt
    \item new PN information, in particular of the spin-shifted dynamical tides~\cite{Steinhoff:2021dsn}, 7.5PN order adiabatic effects in the waveform modes~\cite{Henry:2020ski,Mandal:2024,Dones:2024}, and 3.5PN spin-induced multipole moments~\cite{Henry:2022dzx,Khalilv5},
    \item the development of a pre-merger model and calibration to NR waveforms across a wide range of tidal deformabilities with mass ratios up to $q = 2$,
    \item the higher modes,
    \item drastically improved speed for $M \geq 2\,M_\odot$ systems by a factor 100 to 1000, making it fast enough to be used in PE for current detectors, and
    \item the state-of-the-art BBH model \vfivehm~\cite{Pompiliv5} as the BBH baseline.
\end{enumerate}

In summary, the model exhibits enhancements in both speed and accuracy when compared to the prior generation model \vfourt.

To validate \vfivet\, we study its dephasing and mismatch relative to existing NR waveforms, its predecessor \vfourt, as well as the other state-of-the-art BNS waveform models \teobg, \imrphenomxpnrtidalthree\ and \seobnrvfiveromnrtidalvthree. Our findings reveal that, while all waveform models examined in this study perform similarly in predicting waveforms from BAM simulations, only \vfivet\ generates waveforms that are of the same order of error as the intrinsic NR uncertainty for both BAM and SACRA simulations. A noteworthy caveat of this result is, however, that although the SACRA simulations were generated at a higher resolutions than the BAM simulations, they, in contrast to the BAM simulations, do not converge at a clear convergence order. Our error estimate for the BAM simulations is therefore more robust than our error estimate for the SACRA simulations, which might be underestimated.

A comparison of the current-generation BNS waveform approximants across the relevant BNS parameter space reveals noticeable differences amongst the waveforms for spin magnitudes $|\chi|\gtrsim 0.3$, high primary tidal deformabilities $\Lambda_2^{(1)}\gtrsim 4000$, and high total masses $M\gtrsim 6\ M_\odot$. This is to be expected due to the limited number of NR simulations in these parts of the parameter space. Furthermore, while the EOB BNS models appear to agree for mass-ratios up to $q=6$, noticeable differences between \vfivet\ and \nrtidal-variants appear for large mass-ratios $q\gtrsim 2$.

As \vfivet\ is however the only current model that includes the shift in the $f$-mode frequency due to spin, we believe that it should be the most accurate spin-aligned BNS model for small to medium spins $|\chi| \lesssim 0.3$ in comparison to the other available approximants, and that it can be employed for masses $m_{A,B}\in [0.5,3]\ M_\odot$, and for dimensionless tidal deformabilities $\Lambda_{A,B}\in [0,5000]$, although the model is robust enough to generate waveforms also for higher mass-ratios and highly spinning BNS systems, if one starts the evolution from a low enough initial frequency.

\begin{figure*}[t!]
    \centering 
  \includegraphics[width=0.95\linewidth]{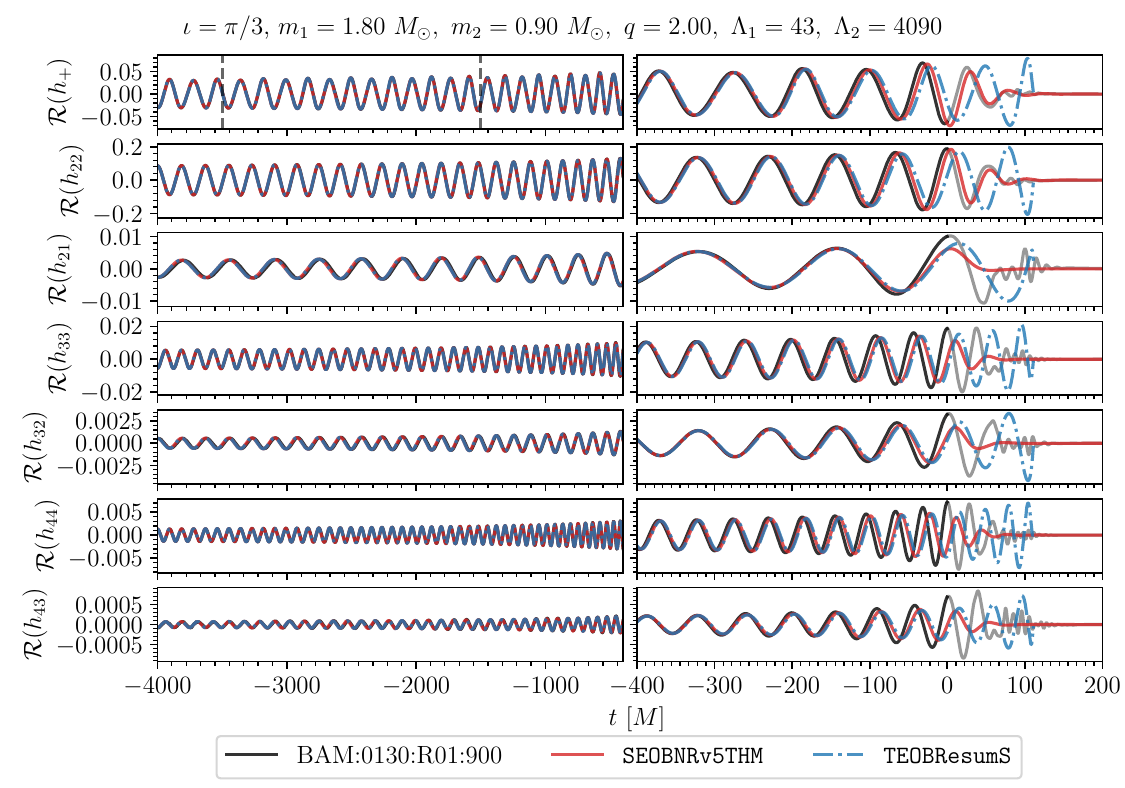}
  \includegraphics[width=0.95\linewidth]{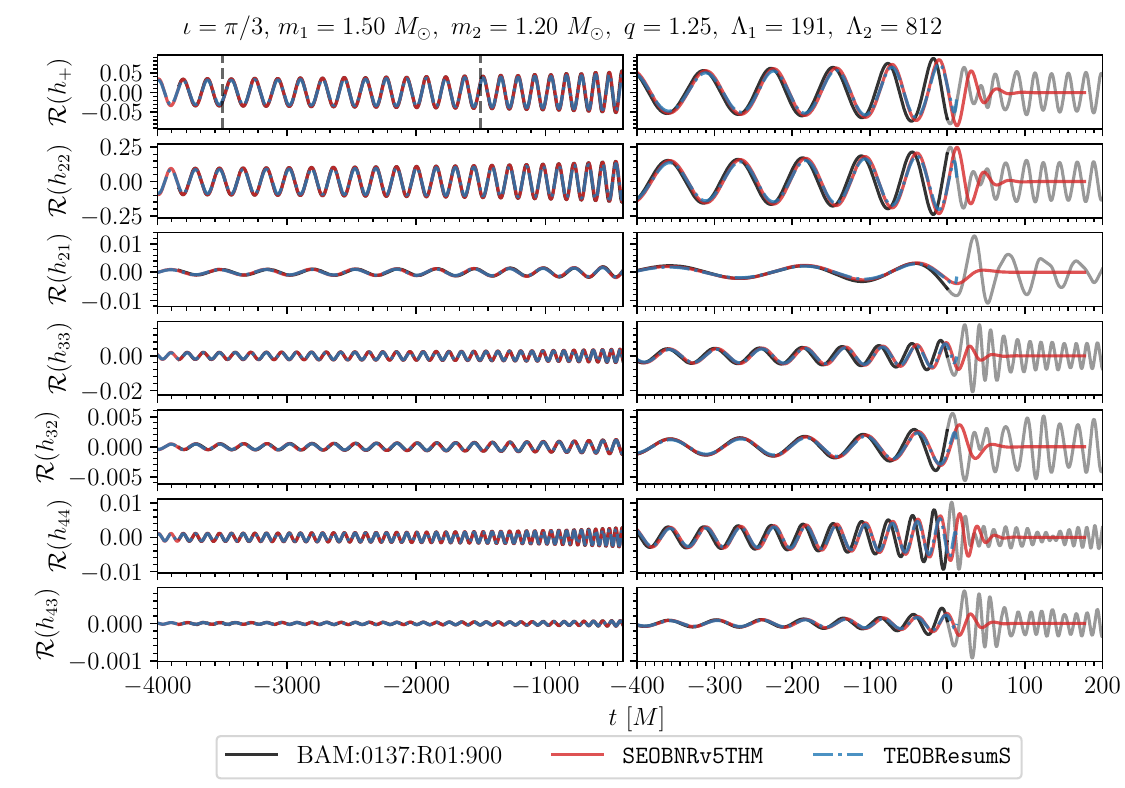}
\caption{Comparison between the GW polarization and modes of \vfivet\ and \teob\ against NR simulations (black), where the post-merger NR waveform is shown in grey. The system parameters are stated above each figure and we depict the real part of the + polarization $h_+$ in the first row, and the higher modes as aligned through the polarization in the lower rows. The vertical dashed lines mark the time window where the alignment against $h_+^{\rm NR}$ is done. Note that \teob\ has no tapering prescription in time-domain and ends abruptly.}
\label{fig:hm}
\end{figure*}

Suitable future developments for \vfivet\ would be in order of presumed feasibility:
\begin{enumerate}[label=(\roman*)]
    \itemsep-3pt
    \item the inclusion of the LO adiabatic tidal effects for $l\geq 4$, gravitomagnetic tidal effects, and the respective tidal adjustment of the $\bar{D}_\text{noS}$-potential as is done in \teob~\cite{Akcay:2018yyh,Gamba:2024cvy};
    \item the further calibration and validation to new BNS and black-hole-neutron-star simulations~\cite{Markin:2023fxx,LIGOScientific:2021qlt,Kuan:2024};
    \item the accurate modeling of the pre-merger signal of the higher-order modes, as \vfivet\ only calibrates the (2,2) mode during the pre-merger;
    \item the application of the stationary-phase-approximation when Fourier transforming~\cite{Gamba:2020ljo};
    \item the continued study of resonance phenomena for highly spinning, anti-aligned NSs through analytical and numerical advances~\cite{Yu:2024,Kuan:2024}, including the tidal spin~\cite{Steinhoff:2016rfi,Steinhoff:2021dsn,Yu:2025};
    \item the addition of precession and eccentric effects on top of the aligned-spin model; and
    \item the development of a full theoretical understanding of dynamical tides~\cite{Mandal:2023hqa}, in particular for precessing and eccentric BNS systems.
\end{enumerate}

We believe that \vfivet\ is an excellent starting point for such future research.

\section*{\label{section: Acknowledgments}Acknowledgments}

It is our pleasure to thank Adrian Abac for sharing his code, his PE results for \imrphenomxpnrtidalthree, and his knowledge related to $(2,2)$ mode phase-alignments, NR simulation access, and PE with us, as well as for the many helpful discussion. We would furthermore like to thank 
Aurora Abbondanza,
Marta Colleoni,
Tim Dietrich,
Raffi Enficiaud,
Hector Estelles,
Guglielmo Faggioli,
Carla Ferradini,
Cheng Foo,
Aldo Gamboa,
Nihar Gupte,
Quentin Henry,
Mohammed Khalil,
Saketh Maddu,
Peter James Nee,
Nami Nishimura,
Serguei Ossokine,
Raj Patil,
Lorenzo Pompili,
Hector Silva,
Maarten van de Meent, and
Sebastian V\"olkel
for helpful discussions that have improved this work.

This material is based upon work supported by NSF's LIGO Laboratory which is a major facility fully funded by the National Science Foundation. The computational work for this manuscript was carried out on the compute cluster Hypatia at
the Max Planck Institute for Gravitational Physics in Potsdam.

\begin{table*}[t]
    \caption{\label{tab:nr_BAM} Properties of the 64 NR waveforms used for \vfivet. For each waveform, we indicate the EoS, the individual masses $m_{1,2}\ [M_{\odot}]$ of the individual bodies, the mass ratio $q$, the dimensionless tidal deformabilities $\Lambda_{2}^{(1,2)}$, the radii $R_{1,2}$ [km] and dimensionless spins $\chi_{1,2}$.}
\begin{ruledtabular}
    \begin{tabular}{l || cccccccccc}
    \textbf{NR Waveform Name}             & EoS  & $m_1$ & $m_2$ & $q$  & $\Lambda_{2}^{(1)}$ & $\Lambda_{2}^{(2)}$ & $R_1$ & $R_2$ & $\chi_1$ & $\chi_2$ \\ \hline \hline
    CoRe:BAM:0001 & 2B & 1.35 & 1.35 & 1.00 & 127 & 127 & 9.72 & 9.72 & 0 & 0 \\ 
    CoRe:BAM:0037 & H4 & 1.37 & 1.37 & 1.00 & 1006 & 1006 & 13.54 & 13.54 & 0 & 0 \\ 
    CoRe:BAM:0039 & H4 & 1.37 & 1.37 & 1.00 & 1002 & 1002 & 13.54 & 13.54 & 0.27 & 0.27 \\ 
    CoRe:BAM:0039 & H4 & 1.37 & 1.37 & 1.00 & 1002 & 1002 & 13.54 & 13.54 & 0.27 & 0.27 \\ 
    CoRe:BAM:0062 & MS1b & 1.35 & 1.35 & 1.00 & 1532 & 1532 & 14.02 & 14.02 & -0.18 & -0.18 \\ 
    CoRe:BAM:0064 & MS1b & 1.35 & 1.35 & 1.00 & 1532 & 1532 & 14.01 & 14.01 & 0 & 0 \\ 
    CoRe:BAM:0066 & MS1b & 1.35 & 1.35 & 1.00 & 1532 & 1532 & 14.02 & 14.02 & 0.18 & 0.18 \\ 
    CoRe:BAM:0068 & MS1b & 1.35 & 1.35 & 1.00 & 1525 & 1525 & 14.01 & 14.01 & 0.27 & 0.27 \\ 
    CoRe:BAM:0095 & SLy & 1.35 & 1.35 & 1.00 & 390 & 390 & 11.46 & 11.46 & 0 & 0 \\ 
    CoRe:BAM:0096 & SLy & 1.35 & 1.35 & 1.00 & 390 & 390 & 11.46 & 11.46 & 0 & 0 \\ 
    CoRe:BAM:0097 & SLy & 1.35 & 1.35 & 1.00 & 390 & 390 & 11.46 & 11.46 & 0 & 0 \\ 
    CoRe:BAM:0101 & SLy & 1.35 & 1.35 & 1.00 & 390 & 390 & 11.46 & 11.46 & 0.10 & 0.10 \\ 
    CoRe:BAM:0104 & SLy & 1.35 & 1.35 & 1.00 & 388 & 388 & 11.46 & 11.46 & 0.19 & 0.19 \\ 
    CoRe:BAM:0129 & SLy & 1.35 & 1.35 & 1.00 & 390 & 390 & 11.47 & 11.47 & 0.17 & 0.17 \\ 
    CoRe:BAM:0130 & SLy & 1.80 & 0.90 & 2.00 & 43 & 4090 & 11.14 & 11.36 & 0 & 0 \\ 
    CoRe:BAM:0131 & SLy & 1.72 & 0.98 & 1.75 & 66 & 2535 & 11.27 & 11.39 & 0 & 0 \\ 
    CoRe:BAM:0132 & SLy & 1.35 & 1.35 & 1.00 & 390 & 390 & 11.46 & 11.46 & 0 & 0 \\ 
    CoRe:BAM:0136 & SLy & 1.62 & 1.08 & 1.50 & 108 & 1489 & 11.35 & 11.42 & 0 & 0 \\ 
    CoRe:BAM:0137 & SLy & 1.50 & 1.20 & 1.25 & 191 & 812 & 11.42 & 11.45 & 0 & 0 \\ 
    CoRe:BAM:0141 & MS1b & 1.40 & 1.37 & 1.02 & 1260 & 1416 & 14.06 & 14.03 & 0.21 & 0.08 \\ 
    CoRe:BAM:0144 & SLy & 1.35 & 1.35 & 1.00 & 390 & 390 & 11.47 & 11.47 & -0.17 & -0.17 \\
        \end{tabular}
    \end{ruledtabular}
\end{table*}

\begin{table*}[t] 
    \caption{\label{tab:nr_SACRA} Properties of the 64 NR waveforms used for \vfivet. For each waveform, we indicate the EoS, the individual masses $m_{1,2}$ of the individual bodies, the mass ratio $q$, the dimensionless tidal deformabilities $\Lambda_{2}^{(1,2)}$, the radii $R_{1,2}$ and dimensionless spins $\chi_{1,2}$.}
\begin{ruledtabular}
    \begin{tabular}{l || cccccccccc}
    \textbf{NR Waveform Name}             & EoS  & $m_1\ [M_{\odot}]$ & $m_2\ [M_{\odot}]$ & $q$  & $\Lambda_{2}^{(1)}$ & $\Lambda_{2}^{(2)}$ & $R_1$ [km] & $R_2$ [km] & $\chi_1$ & $\chi_2$ \\ \hline \hline
    SACRA:15H\_135\_135\_00155\_182\_135  & 15H  & 1.35  & 1.35  & 1.00 & 1211                                                       & 1211                                                       & 13.69 & 13.69 & 0        & 0        \\
    SACRA:125H\_135\_135\_00155\_182\_135 & 125H & 1.35  & 1.35  & 1.00 & 863                                                        & 863                                                        & 12.97 & 12.97 & 0        & 0        \\
    SACRA:H\_135\_135\_00155\_182\_135    & H    & 1.35  & 1.35  & 1.00 & 607                                                        & 607                                                        & 12.27 & 12.27 & 0        & 0        \\
    SACRA:HB\_135\_135\_00155\_182\_135   & HB   & 1.35  & 1.35  & 1.00 & 422                                                        & 422                                                        & 11.61 & 11.61 & 0        & 0        \\
    SACRA:B\_135\_135\_00155\_182\_135    & B    & 1.35  & 1.35  & 1.00 & 289                                                        & 289                                                        & 10.96 & 10.96 & 0        & 0       \\
    SACRA:15H\_125\_146\_00155\_182\_135  & 15H  & 1.46  & 1.25  & 1.17 & 760                                                        & 1871                                                       & 13.72 & 13.65 & 0        & 0        \\
    SACRA:125H\_125\_146\_00155\_182\_135 & 125H & 1.46  & 1.25  & 1.17 & 535                                                        & 1351                                                       & 12.99 & 12.94 & 0        & 0        \\
    SACRA:H\_125\_146\_00155\_182\_135    & H    & 1.46  & 1.25  & 1.17 & 369                                                        & 966                                                        & 12.18 & 12.26 & 0        & 0        \\
    SACRA:HB\_125\_146\_00155\_182\_135   & HB   & 1.46  & 1.25  & 1.17 & 252                                                        & 684                                                        & 11.59 & 11.61 & 0        & 0        \\
    SACRA:15H\_121\_151\_00155\_182\_135  & 15H  & 1.51  & 1.21  & 1.25 & 625                                                        & 2238                                                       & 13.73 & 13.63 & 0        & 0        \\
    SACRA:125H\_121\_151\_00155\_182\_135 & 125H & 1.51  & 1.21  & 1.25 & 435                                                        & 1621                                                       & 12.98 & 12.93 & 0        & 0        \\
    SACRA:H\_121\_151\_00155\_182\_135    & H    & 1.51  & 1.21  & 1.25 & 298                                                        & 1163                                                       & 12.26 & 12.25 & 0        & 0        \\
    SACRA:HB\_121\_151\_00155\_182\_135   & HB   & 1.51  & 1.21  & 1.25 & 200                                                        & 827                                                        & 11.57 & 11.60 & 0        & 0        \\
    SACRA:B\_121\_151\_00155\_182\_135    & B    & 1.51  & 1.21  & 1.25 & 131                                                        & 581                                                        & 10.89 & 10.98 & 0        & 0        \\
    SACRA:15H\_118\_155\_00155\_182\_135  & 15H  & 1.55  & 1.18  & 1.31 & 530                                                        & 2575                                                       & 13.74 & 13.62 & 0        & 0        \\
    SACRA:125H\_118\_155\_00155\_182\_135 & 125H & 1.55  & 1.18  & 1.31 & 366                                                        & 1875                                                       & 12.98 & 12.92 & 0        & 0        \\
    SACRA:H\_118\_155\_00155\_182\_135    & H    & 1.55  & 1.18  & 1.31 & 249                                                        & 1354                                                       & 12.26 & 12.24 & 0        & 0        \\
    SACRA:HB\_118\_155\_00155\_182\_135   & HB   & 1.55  & 1.18  & 1.31 & 165                                                        & 966                                                        & 11.55 & 11.60 & 0        & 0        \\
    SACRA:B\_118\_155\_00155\_182\_135    & B    & 1.55  & 1.18  & 1.31 & 107                                                        & 681                                                        & 10.87 & 10.98 & 0        & 0        \\
    SACRA:15H\_117\_156\_00155\_182\_135  & 15H  & 1.56  & 1.17  & 1.33 & 509                                                        & 2692                                                       & 13.74 & 13.61 & 0        & 0        \\
    SACRA:125H\_117\_156\_00155\_182\_135 & 125H & 1.56  & 1.17  & 1.33 & 350                                                        & 1963                                                       & 12.98 & 12.91 & 0        & 0        \\
    SACRA:H\_117\_156\_00155\_182\_135    & H    & 1.56  & 1.17  & 1.33 & 238                                                        & 1415                                                       & 12.25 & 12.24 & 0        & 0        \\
    SACRA:HB\_117\_156\_00155\_182\_135   & HB   & 1.56  & 1.17  & 1.33 & 157                                                        & 1013                                                       & 11.55 & 11.60 & 0        & 0        \\
    SACRA:B\_117\_156\_00155\_182\_135    & B    & 1.56  & 1.17  & 1.33 & 101                                                        & 719                                                        & 10.86 & 10.98 & 0        & 0        \\
    SACRA:15H\_116\_158\_00155\_182\_135  & 15H  & 1.58  & 1.16  & 1.36 & 465                                                        & 2863                                                       & 13.73 & 13.60 & 0        & 0        \\
    SACRA:125H\_116\_158\_00155\_182\_135 & 125H & 1.58  & 1.16  & 1.36 & 319                                                        & 2085                                                       & 12.98 & 12.90 & 0        & 0        \\
    SACRA:H\_116\_158\_00155\_182\_135    & H    & 1.58  & 1.16  & 1.36 & 215                                                        & 1506                                                       & 12.25 & 12.23 & 0        & 0        \\
    SACRA:HB\_116\_158\_00155\_182\_135   & HB   & 1.58  & 1.16  & 1.36 & 142                                                        & 1079                                                       & 11.53 & 11.59 & 0        & 0        \\
    SACRA:B\_116\_158\_00155\_182\_135    & B    & 1.58  & 1.16  & 1.36 & 91                                                         & 765                                                        & 10.84 & 11.98 & 0        & 0        \\
    SACRA:15H\_125\_125\_0015\_182\_135   & 15H  & 1.25  & 1.25  & 1.00 & 1875                                                       & 1875                                                       & 13.65 & 13.65 & 0        & 0        \\
    SACRA:125H\_125\_125\_0015\_182\_135  & 125H & 1.25  & 1.25  & 1.00 & 1352                                                       & 1352                                                       & 12.94 & 12.94 & 0        & 0        \\
    SACRA:H\_125\_125\_0015\_182\_135     & H    & 1.25  & 1.25  & 1.00 & 966                                                        & 966                                                        & 12.26 & 12.26 & 0        & 0        \\
    SACRA:HB\_125\_125\_0015\_182\_135    & HB   & 1.25  & 1.25  & 1.00 & 683                                                        & 683                                                        & 11.61 & 11.61 & 0        & 0        \\
    SACRA:B\_125\_125\_0015\_182\_135     & B    & 1.25  & 1.25  & 1.00 & 476                                                        & 476                                                        & 10.98 & 10.98 & 0        & 0        \\
    SACRA:15H\_112\_140\_0015\_182\_135   & 15H  & 1.40  & 1.12  & 1.25 & 975                                                        & 3411                                                       & 13.71 & 13.58 & 0        & 0        \\
    SACRA:125H\_112\_140\_0015\_182\_135  & 125H & 1.40  & 1.12  & 1.25 & 693                                                        & 2490                                                       & 12.98 & 12.89 & 0        & 0        \\
    SACRA:H\_112\_140\_0015\_182\_135     & H    & 1.40  & 1.12  & 1.25 & 484                                                        & 1812                                                       & 12.28 & 12.23 & 0        & 0        \\
    SACRA:HB\_112\_140\_0015\_182\_135    & HB   & 1.40  & 1.12  & 1.25 & 333                                                        & 1304                                                       & 11.60 & 11.59 & 0        & 0        \\
    SACRA:B\_112\_140\_0015\_182\_135     & B    & 1.40  & 1.12  & 1.25 & 225                                                        & 933                                                        & 10.95 & 10.97 & 0        & 0        \\
    SACRA:15H\_107\_146\_0015\_182\_135   & 15H  & 1.46  & 1.07  & 1.36 & 760                                                        & 4361                                                       & 13.72 & 13.54 & 0        & 0        \\
    SACRA:125H\_107\_146\_0015\_182\_135  & 125H & 1.46  & 1.07  & 1.36 & 535                                                        & 3196                                                       & 12.99 & 12.86 & 0        & 0        \\
    SACRA:H\_107\_146\_0015\_182\_135     & H    & 1.46  & 1.07  & 1.36 & 369                                                        & 2329                                                       & 12.18 & 12.22 & 0        & 0        \\
    SACRA:HB\_107\_146\_0015\_182\_135    & HB   & 1.46  & 1.07  & 1.36 & 252                                                        & 1695                                                       & 11.59 & 11.60 & 0        & 0        \\
    SACRA:B\_107\_146\_0015\_182\_135     & B    & 1.46  & 1.07  & 1.36 & 168                                                        & 1216                                                       & 10.92 & 10.97 & 0        & 0        \\
    SACRA:SFHo\_135\_135\_00155\_182\_135 & SFHo & 1.35  & 1.35  & 1.00 & 460                                                        & 460                                                        & 11.91 & 11.91 & 0        & 0        \\
\end{tabular}
\end{ruledtabular}
\end{table*}

\appendix

\section{Quasi-universal relations} \label{app:URs}

In this appendix, we collect all the universal relations used in the waveform model. We use the spin-induced multipole moments $C_l$, as a function of $\Lambda_2$, from Refs.~\cite{Yagi:2016bkt,Abac:NRT}, the octupolar tidal deformability $\Lambda_3$ from Ref.~\cite{Yagi:2014bxa}, and $\omega_{0l}$ from Refs.~\cite{Yagi:2014bxa,Sotani:2021}.

To find a good estimate for $\omega_{f,l}(\Lambda, \chi)$ we compute the non-spinning moment of inertia $I$~\cite{Yagi:2016bkt} and the NS compactness and radius~\cite{Gamba:2021prd}.
We then use Ref.~\cite{Kruger:2023} to obtain the spin-adjusted moment of inertia $I(\Lambda, \Omega)$ and invert this numerically to get the spin-angular-frequency $\Omega = I/\chi$.
Finally, we use $\Omega$ in Ref.~\cite{Kruger:2020} to arrive at $\Delta \omega_{0l}$.

\section{Tapering of the subdominant modes} \label{app:tapering}

The tapering of the subdominant $(\ell, |m|)=(3,3),(2,1),(4,4),(5,5),(3,2),$ and $(4,3)$ modes builds heavily on the tapering of \vfourt~\cite{Lackey:2018zvw}, but we change it to satisfy a continuously differentiable frequency evolution as described in Sec.~\ref{subsec:Waveform}.

The steps for the tapering of the inspiral-plunge waveform $h_{\ell m}^{\rm pre-merge} = A_{\ell m} \exp (-i\phi_{\ell m})$ are:
\begin{enumerate}[label=(\roman*)]
    \item linearly continue the intermediate amplitude from the inspiral-plunge amplitude 
    \begin{align}
        \label{eq:Aboost_sub}
        \hat{A}_{\ell m}(t) = \begin{cases}
                A_{\ell m}(t) \qquad\qquad\quad \mathrm{for}\ t \leq t_\mathrm{match}, \\\\
            \begin{aligned}[t]
                & A_{\ell m}(t_\mathrm{match}) \\ &+ \dot{A}_{\ell m} (t_\mathrm{match}) (t-t_\mathrm{match}) \\
                & \qquad\qquad\qquad\qquad\mathrm{for}\ t > t_\mathrm{match};
            \end{aligned}
        \end{cases}
    \end{align}
    \item extract the final waveform phase $\phi_\mathrm{match}^{\ell m} =\phi_{\ell m}(t_\mathrm{match})$ and frequency $\omega^{\ell m}_\mathrm{match} =\omega_{\ell m}(t_\mathrm{match})$ and the respective phase and frequency 12 $M$ prior to $\tmatch$: $\phi^{\ell m}_\mathrm{freq} =\phi_{\ell m}(t_\mathrm{match} - 12\ M)$ and $\omega^{\ell m}_\mathrm{freq} =\omega_{\ell m}(t_\mathrm{match}-12\ M)$;
    \item let the phase transition from $\tmatch - 12\ M$ onwards to the asymptotic frequency $\omega_\mathrm{match}$ through a tapering of the form
    \begin{align}
        \label{eq:freqboost_sub}
        \hat{\omega}_{\ell m}(t) = \begin{cases}
                \omega_{\ell m}(t) \qquad \mathrm{for}\ t \leq t_\mathrm{match} - 12\ M, \\\\
            \begin{aligned}[t]
                & \omega^{\ell m}_\mathrm{match} \\ &- \Delta \omega_{\ell m} \exp \left(-\frac{t-t_\mathrm{match}}{\tau_\mathrm{freq}^{\ell m}}\right) \\
                & \qquad\qquad\quad \mathrm{for}\ t > t_\mathrm{match} - 12\ M,
            \end{aligned}
        \end{cases}
    \end{align}
   where $\Delta \omega_{\ell m} = \omega^{\ell m}_\mathrm{match} - \omega^{\ell m}_\mathrm{freq}$ and $\tau_\mathrm{freq}^{\ell m}$ is chosen to ensure continuity of the first frequency derivative across the whole waveform;
    \item use the Planck taper from Eq.~\eqref{eq:transition} with decay time tuned to be the same as for the dominant $(2,2)$ mode $\tau_W=0.5\pi/\omega_{22}(t_\mathrm{match})$ centred $15\ M$ prior to $\tmatch$
    \begin{equation}
        W(t)=\left\{1 + \exp\left[\frac{t - (t_{\rm match} - 15\ M)}{\tau_W}\right]\right\}^{-1};
    \end{equation}
     \item the tapered waveform then has the form \begin{equation}
        h_{\ell m}(t)=\hat{A}_{\ell m}(t)\, W(t)\,\exp\left(-i\int^t \hat{\omega}_{\ell m}(t')\d t' \right),
     \end{equation} 
     with 
     \begin{equation}
        \begin{split}
            \int^{t} &\hat{\omega}_{\ell m}(t')\d t' = \phi^{\ell m}_\mathrm{match} + \omega^{\ell m}_\mathrm{match} (t-t_\mathrm{freq}) \\ &+ \tau^{\ell m}_\mathrm{freq} \Delta \omega_{\ell m} \left[\exp \left(-\frac{t-t_\mathrm{match}}{\tau^{\ell m}_\mathrm{freq}}\right)-1\right].
        \end{split}
     \end{equation}
    \end{enumerate}

    \section{Validation of the higher modes} \label{app:hm}

In Fig.~\ref{fig:hm} we show the gravitational polarization $h_+=\mathrm{Re}(h)$ from Eq.~\eqref{eq:hoft_sphericalH} as well as the higher modes 
\begin{equation}
    (\ell,\,m)\in\{(2,2),\,(2,1),\,(3,3),\,(3,2),\,(4,4),\,(4,3)\}
\end{equation}
for two of the four publicly available BNS NR simulations with available higher-modes decomposition~\cite{Ujevic:2022qle} at an inclination $\iota=\pi/3$.

To align the modes, we minimize the difference in the frequency-time series of the NR and EOB waveforms
\begin{equation} \label{eq:hm_alignment}
    \min_{(t_c,\,\phi_c)}\ \int_{t_{\rm start}}^{t_{\rm end}} |\omega^{\rm (NR)}_{+}(t,\iota,\phi_c) - \omega^{\rm (EOB)}_{+}(t - t_c,\iota,\phi_c)|^2\, \d t,
\end{equation}
where 
\begin{equation}
    \omega_{\rm +}(t,\iota,\phi_c) = \frac{\d \phi_{\rm +}(t,\iota,\phi_c)}{\d t}
\end{equation}
is the angular frequency of the + polarization
\begin{equation}
    h_+ = A_+(t,\iota,\phi_c) e^{i \phi_{\rm +}(t,\iota,\phi_c)}.
\end{equation}
We then transform each GW mode by the obtained azimuthal angle
\begin{equation}
    \phi_{\ell m}(t)\mapsto \phi_{\ell m}(t-t_c)e^{im\phi_c}.
\end{equation}

We note that we only show comparisons to the highest-resolution simulations. We find that we introduce artifacts into the waveform phase when performing a Richardson interpolation towards merger, and that the alignment against the polarization becomes harder and tends to not align the higher modes well, which we leave to be studied more in-depth for the future.

We make the observation that for near equal-mass systems $q\sim 1$, both \teob\ and \vfivet\ agree very well with each other, while for higher mass ratios $q\sim 2$ \vfivet\ performs better than \teob\ due to the more accurate time of merger across all modes, which remains true also for the two simulations not shown in this work. We remind the reader that the time-domain waveform of \teob\ has no tapering prescription and therefore ends abruptly at the merger.

\section{NR simulations used in this work}
\label{app:NR_calib}

In Table~\ref{tab:nr_BAM} and Table~\ref{tab:nr_SACRA} we show the NS parameters of all the 64 NR waveforms from SACRA~\cite{Kiuchi:2017pte, Kawaguchi:2018gvj, Kiuchi:2019kzt} and CoRe (using the BAM code)~\cite{Dietrich:2018phi, Ujevic:2022qle, Gonzalez:2022mgo} that were used in this work partially in the calibration of the merger time and fully in the validation of \vfivet\ in Sec.~\ref{sec:Validation}. We list the individual masses of the stars $m_{1,2}$, the mass ratio $q$, dimensionless tidal deformabilities $\Lambda_2^{(1,2)}$, and radii $R_{1,2}$. The radii data for the SACRA waveforms were adapted from Ref.~\cite{Kiuchi:2017pte, Kawaguchi:2018gvj, Kiuchi:2019kzt}, while we compute the radii for the CORE (BAM) waveforms using the adiabatic Love number and tidal deformability, i.e. from $\Lambda_2^{(i)} = (2/3)k_2^{(i)}R_i^5/m_i^5$.


\bibliography{Bibliography}

\end{document}